\documentclass[twocolumn]{aastex631}

\usepackage{commath}
\usepackage{floatrow}
\usepackage{soul}
\usepackage{CJK}
\usepackage{threeparttable}
\usepackage{makecell}
\usepackage{multirow}
\usepackage{txfonts}

\newcommand{\subsubsubsection}[1]{\paragraph{#1}\mbox{}\\}
\shorttitle{\textit{ Chocolate Chip Cookie} Model}
\shortauthors{Lu et al.}
\received{}
\revised{}
\accepted{}
\submitjournal{ApJ}

\graphicspath{{./}{figures/}}

\begin{document}
\begin{CJK*}{UTF8}{gbsn}

\title{The\textit{ Chocolate Chip Cookie} Model:  Dust Geometry of Milky-Way like Disk Galaxies}

\correspondingauthor{Shiyin Shen}
\email{ssy@shao.ac.cn}

\author[0000-0002-8817-4587]{Jiafeng Lu (卢家风)}
\affiliation{Key Laboratory for Research in Galaxies and Cosmology, Shanghai Astronomical Observatory, Chinese Academy of Sciences, 80 Nandan Road, Shanghai 200030, People's Republic of China}
\affiliation{University of Chinese Academy of Sciences, 19A Yuquan Road, Beijing 100049, People's Republic of China}

\author[0000-0002-3073-5871]{Shiyin Shen (沈世银)}
\affiliation{Key Laboratory for Research in Galaxies and Cosmology, Shanghai Astronomical Observatory, Chinese Academy of Sciences, 80 Nandan Road, Shanghai 200030, People's Republic of China}
\affiliation{Key Lab for Astrophysics, Shanghai, 200034, People's Republic of China}

\author[0000-0001-6763-5869]{Fang-Ting Yuan (袁方婷)}
\affiliation{Key Laboratory for Research in Galaxies and Cosmology, Shanghai Astronomical Observatory, Chinese Academy of Sciences, 80 Nandan Road, Shanghai 200030, People's Republic of China}
\affiliation{Key Lab for Astrophysics, Shanghai, 200034, People's Republic of China}

\author[0000-0001-8611-2465]{Zhengyi Shao (邵正义)}
\affiliation{Key Laboratory for Research in Galaxies and Cosmology, Shanghai Astronomical Observatory, Chinese Academy of Sciences, 80 Nandan Road, Shanghai 200030, People's Republic of China}
\affiliation{Key Lab for Astrophysics, Shanghai, 200034, People's Republic of China}

\author{Jinliang Hou (侯金良)}
\affiliation{Key Laboratory for Research in Galaxies and Cosmology, Shanghai Astronomical Observatory, Chinese Academy of Sciences, 80 Nandan Road, Shanghai 200030, People's Republic of China}
\affiliation{University of Chinese Academy of Sciences, 19A Yuquan Road, Beijing 100049, People's Republic of China}

\author[0000-0003-3728-9912]{Xianzhong Zheng (郑宪忠)}
\affiliation{Purple Mountain Observatory, Chinese Academy of Sciences, 8 Yuan Hua Road, Nanjing, Jiangsu 210033, People's Republic of China}

\begin{abstract}

We present a new two-component dust geometry model,  the \textit{Chocolate Chip Cookie} model,  where the clumpy nebular regions are embedded in a diffuse stellar/ISM disk, like chocolate chips in a cookie.  By approximating the binomial distribution of the clumpy nebular regions with a continuous  Gaussian distribution and omitting the dust scattering effect,  our model solves the dust attenuation process for both the emission lines and stellar continua via analytical approaches. Our \textit{Chocolate Chip Cookie} model successfully fits the inclination dependence of both the effective dust reddening of the stellar components derived from stellar population synthesis and that of the emission lines characterized by the Balmer decrement for a large sample of Milky-Way like disk galaxies selected from the main galaxy sample of the Sloan Digital Sky Survey (SDSS).  Our model shows that the clumpy nebular disk is about 0.55 times thinner and 1.6 times larger than the stellar disk for MW-like galaxies, whereas each clumpy region has a typical optical depth $\tau_{\rm{cl,V}} \sim 0.5$ in $V$ band. After considering the aperture effect, our model prediction on the inclination dependence of dust attenuation is also consistent with observations. Not only that, in our model, the dust attenuation curve of the stellar population naturally  depends on inclination and its median case is consistent with the classical Calzetti law. Since the modelling constraints are from the optical wavelengths, our model is unaffected by the optically thick dust component, which however could bias the model's prediction of the infrared emissions. 
\end{abstract}

\keywords{Disk galaxies(391)-Galaxy structure(622)-Extinction(505)-Interstellar dust(836)-Interstellar dust extinction(837)}

\section{Introduction}
\label{introduction}
Dust accounts for only a tiny fraction ($\sim0.1\%$) of the baryonic mass in star-forming galaxies (SFG) but plays a crucial role in many aspects of galaxy evolution and observational properties. Dust grains are formed in the ejected material of supernova \citep{Ferrarotti2006} and the stellar wind of low-mass stars at the end of their lives \citep{Indebetouw2014, Dwek2011}. They grow, coagulate, and are destructed in the interstellar medium (ISM).  

Dust particles absorb and scatter ultraviolet (UV) and optical photons and re-emit the energy in the infrared (IR) wavelengths. The absorption and scattering of photons lead to a reduction of the emitted flux, known as attenuation, and also cause the reddening of the spectral energy distributions (SED)  since the photons with shorter wavelengths are generally more attenuated than those with longer wavelengths. Because of the attenuation and reddening, dust attenuation is a critical effect to be quantified in deriving the physical properties, such as stellar mass and star formation rate (SFR) of galaxies \citep{Kennicutt1998, Popescu2011, Gadotti2010,Pastrav2013a, Pastrav2013b}.

In the optical spectral analysis, the reddening features of nebular emission lines and stellar continua are the two most commonly used probes of the dust attenuation of a galaxy. At low redshift, Balmer decrement is commonly used to calculate the reddening of the nebular emission, $E_{\rm{g}}(B-V)$. For stellar continuum, the dust reddening $E_{\rm{s}}(B-V)$ can be reliably deduced by using stellar population synthesis (SPS) models.
 
Many studies have shown that the nebular emission in galaxies is more attenuated than the stellar continua \citep{Calzetti1994,Mayya1996,CF00}, which is usually explained by a two-component dust model \citep[e.g.][]{Calzetti2000, CF00}. In this model, the dust distribution in the galaxy has two components: a diffuse ISM component and a clumpy birth-cloud component. Thus, the nebular emission originated from the star-forming regions suffers from the dust attenuation of both the envelope of birth clouds and the diffuse ISM, while the stellar emission, especially that of old stellar populations, most of which is radiated outside of the birth clouds, is extincted only by the diffuse ISM component. Based on the observations of a sample of local starburst galaxies, \citet{Calzetti1997} proposed a typical ratio of the dust reddening for the stars to that for the nebular lines,
\begin{equation}
f\equiv E_{\rm{s}}(B-V)/E_{\rm{g}}(B-V)\sim 0.44,
\end{equation}
which has been widely adopted in many later studies \citep{Madau2014, Peng2010, Daddi2007}. 

However, many recent studies have shown that this stellar-to-nebular dust attenuation ratio $f$ varies systematically with the physical properties of galaxies. For example, \citet{Wild2011} find that $f$ decreases with decreasing specific star formation rate (sSFR) for star-forming galaxies. \citet{Koyama2015} also show that more massive galaxies tend to have higher extra attenuation towards nebular regions, i.e., smaller $f$ values \citep[see also][]{Zahid2017}. {These trends are consistent with the fact that the contribution of old stellar populations, which are less obscured than the young stellar populations that are more closely associated with nebular emission, increases in lower sSFR (or more massive) galaxies.}

The dust attenuation of a galaxy is determined by the contents of its emission sources and dust particles as well as their geometry configuration. As a result, the dust attenuation features of disk galaxies have a strong dependence on their viewing angles. 
Many studies have examined the inclination dependence of the observed properties of disk galaxies \citep[e.g.][]{Shao2007, Maller2009, Masters2010, Yip2010, LiNiu2021, Yuan2021}, and reported that the stellar attenuation is proportional to the logarithm of the observed disk axis ratio $(A_\lambda\propto -\gamma\log (b/a))$ with $\gamma \sim 1$ \citep{Yip2010, Chevallard2013, Battisti2017}, while the nebular attenuation indicated by the Balmer decrements only shows a slight increase trend at low axis ratio. Moreover, \citet{Wild2011} shows that more inclined disks have smaller $f$ values and flatter attenuation curves. On the other hand, these observational trends with disk inclination provide strong constraints on the geometry configuration of disk galaxies. For example, \citet{Yip2010}  conclude that the inclination dependence of the stellar continuum attenuation features deviate significantly from the simple dust screen model,  while the slab model (stars and dust mixed uniformly) and the sandwich model (a layer of dust+stars mixture sandwiched in-between two layers of stars) also cannot account for the observed features appropriately.

Many studies have used radiative transfer (RT) models to constrain the geometry parameters of the dust and stellar components for individual nearby disk galaxies\citep[e.g.,][]{Xilouris1999,Misiriotis2001, Bianchi2007, DeGeyter2014}. \citet{Popescu2000} proposed a multi-component dust distribution model for edge-on disk galaxy NGC 891, which is composed of a bulge, a thick disk associated with diffuse dust and old stellar populations, and a thin disk associated with clumpy dust and newly-formed stars. \citet{tuffs2004} (hereafter T04) extended the multi-component of \citet{Popescu2000} and  further refined the prescription for attenuation of clumpy dust component associated with star-forming regions in the thin disk. For the UV photons emitted from star forming regions, the T04 model assumes that they are either completely absorbed or are able to escape. The fraction of the UV photons being absorbed is then quantified by a free model parameter, clumpiness $F$. With this setting, the T04 model can further make a self-consistent determination of the attenuation of the nebular lines. However,  it is worth pointing out that the clumpiness factor $F$ in  T04 is not a  geometry parameter, where the star-forming regions are continuously distributed.

The T04 model has been widely used to study various dust attenuation properties of disk galaxies \citep{Driver2008,Leslie2018a,Leslie2018b}. For example, by properly setting few model parameters (e.g. $F$ and the disk face-on dust opacity), the T04 model reproduce the observed inclination dependence of the broad band dust attenuation for both low and intermediate redshift galaxies successfully \citep{Driver2007,Masters2010}.  However,  a recent study of \citet{Giessen2022} (hereafter G22) finds that a completely optically thick dust component, as assumed in the T04 model,  can not reproduce the observed inclination-dependence of the Balmer decrement. An extra component of optically thin dust within the birth clouds is required.  In addition, it is worth noting that the attenuation properties derived from T04 model are integrated for global galaxy, while the observed Balmer decrement of SDSS is typically measured from fiber spectroscopy. This spatial mismatch will introduce extra bias into the model that tries to fit these two different attenuation features simultaneously \citep{Chevallard2013}.

In this study, we aim to build a new geometry model of disk galaxies to account for the inclination dependence of the nebular and continuum dust attenuation simultaneously. To avoid of the mismatch of the scales, we  derive both of the nebular and continuum dust attenuation from the fiber spectroscopy. Correspondingly, our modeling of the dust attenuation effects will be mainly along the line of sight to galaxy center rather than make a global estimation. In particular, we will provide a refined modelling of the dust attenuation of the star-forming regions, where its self-extinction and clumpy distribution (as outlined by \citet{CF00}) are both quantified by model parameters. As a starting point of this new model, we will only probe the dust attenuation and reddening features in  optical wavelengths, whereas the dust emission in IR bands will be left for subsequent studies.

The outline of this paper is as follows. In Section \ref{DATA}, we introduce the local star-forming galaxy sample and show the inclination dependence of two different dust reddening features. In Section \ref{simple model}, we model the inclination dependence features with two simple geometric models. In Section \ref{CCC model}, we present the \textit{ Chocolate Chip Cookie} model, a novel two dust component model, and use it to fit the observed inclination dependence of two dust reddening features simultaneously. In Section \ref{APERTURE EFFECT AND DUST ATTENUATION}, we discuss the fiber aperture effect in our sample galaxies and use it to explore the inclination dependence of the emission-line fluxes. In Section \ref{DISCUSSION}, we make further discussions on the best fits of our \textit{Chocolate Chip Cookie } model. Finally, we summarize our main results in Section \ref{CONCLUSION}.

Through this work we adopt the standard cosmology ($H_{\rm{0}}$, $\Omega_{\rm{m}}$, $\Omega_{\rm{\Lambda}}$) $=$ (70\,km\,s$^{-1}$\,Mpc$^{-1}$, 0.3, 0.7).

\section{SAMPLE and DATA}
\label{DATA}
    \subsection{Milky-Way Like Galaxy Sample}
   
Our galaxy sample is drawn from the Main Galaxy Sample (MGS) of the Sloan Digital Sky Survey Data Release 7 (SDSS DR7)\footnote{http://skyserver.sdss.org/dr7/}.  We take the measurements of total stellar mass, fiber magnitude, H$\alpha$, H$\beta$, [OIII]$\lambda$5007 and [NII]$\lambda$6584 emission line fluxes from the MPA-JHU data release \footnote{https://wwwmpa.mpa-garching.mpg.de/SDSS/DR7/} \citep{Kauffmann2003a, Tremonti2004, Brinchmann2004}. To study the dust attenuation properties of both the stellar continua and nebular emission lines, we select star-forming galaxies (SFGs) with emission line features and then use the classical BPT diagram \citep*{Baldwin1981} to remove galaxies with signs of active galactic nuclei (AGNs). We adopt the criteria of \citet{Kauffmann2003b} and require signal-to-noise ratio (S/N) larger than 3 for the four emission lines used. This selection results in 260,856 SFGs.

\begin{figure}[htbp]
\centering
\gridline{\fig{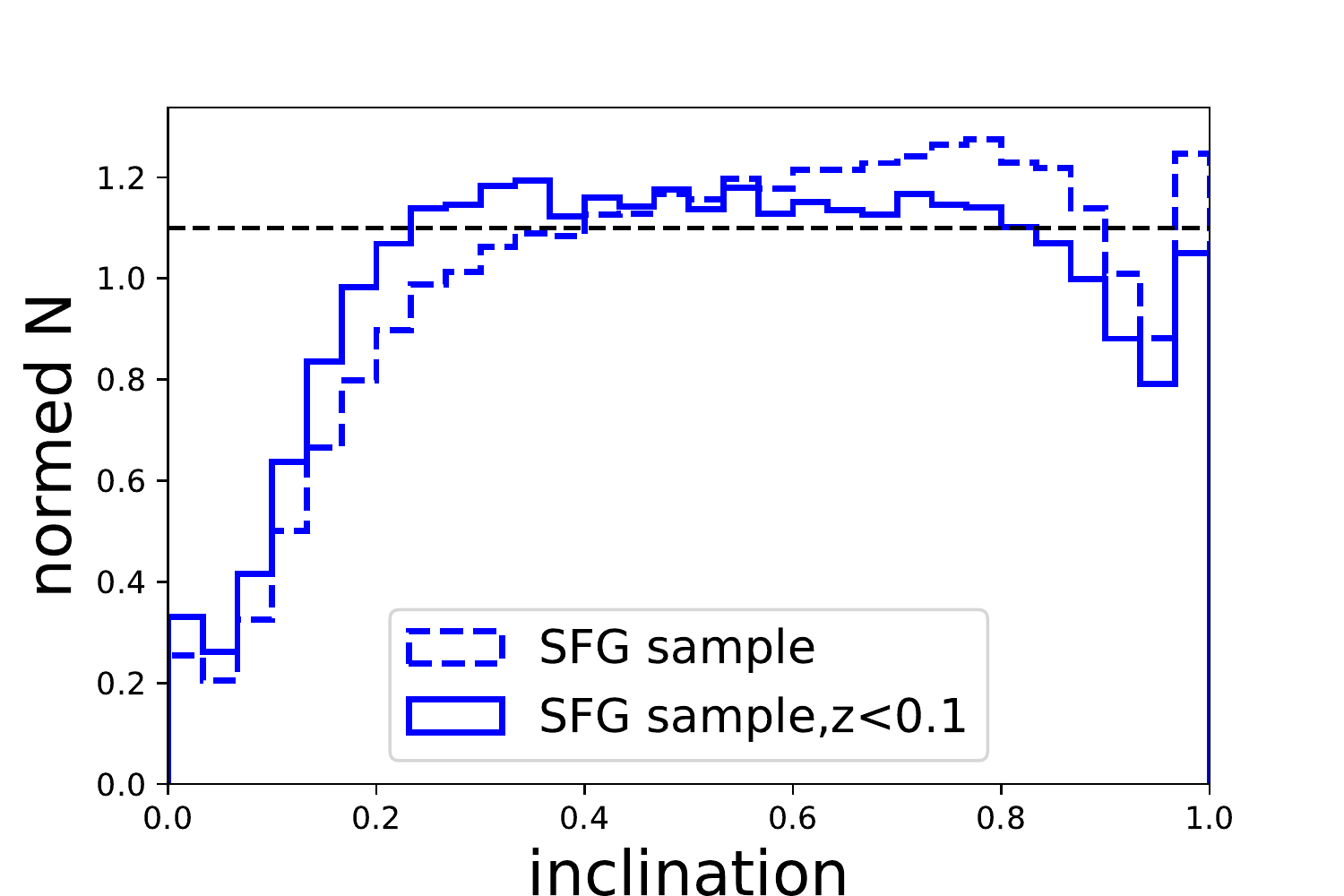}{3in}{(b) histograms of disk inclination}
          }
\gridline{\fig{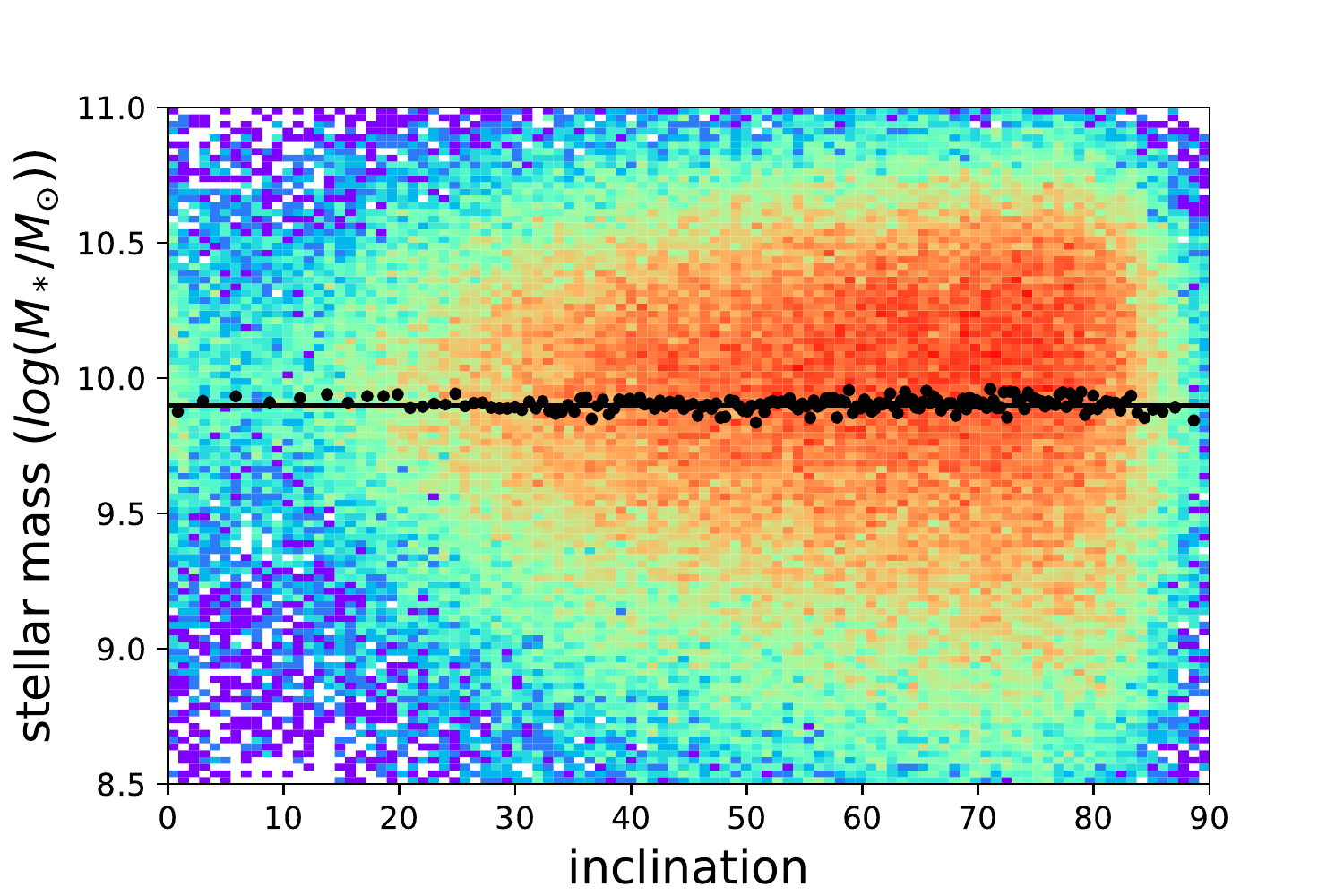}{3in}{(a) mass-inclination dependence}
          }

\caption{ (a) The histograms of disk inclinations.  The dotted line shows  all the SFGs in MGS, while the sold line represents SFGs at  $z<0.1$. Both histograms are normalized to unit area.(b)  The density map  of the stellar mass and disk inclination of the SFGs at $z<0.1$, where the black dots represent median stellar mass at different inclinations.}
\label{incdis}
\end{figure}

We take the stellar mass estimation  from MPA-JHU  and obtain the inclination angle $\theta$ of each SFG from \citet{Simard2011}. According to \citet{Simard2011}, $\theta$ is estimated from morphological fitting by applying a disk+bulge model and therefore gives a better approximation of the real inclination of disk galaxies than the simple axis ratio ($b/a$) parameter, especially for highly inclined disks.  To avoid possible observational biases in the physical properties of our sample galaxies\citep{Driver2007} , we further require sample galaxies at $z<0.1$, where the PSF effects on the modelling of disk inclination could be minimized.   To illustrate this effect, we show the $\cos\theta$ distributions of  all the SFGs in MGS and the selected SFGs at $z<0.1$  as  the dotted and solid  histograms  in the top panel of Figure \ref{incdis} respectively . As can be seen, compared with all SFGs,  SFGs at $z<0.1$ show a much flatter cos$\theta$ distribution,   indicating  a better and unbiased measurement of disk inclination.    In the bottom panel of Figure \ref{incdis}, we show the number density of   the SFGs at $z<0.1$  as function of  their stellar mass and disk inclination.  We see little systematic bias in the stellar mass of these SFGs  at different inclinations.

In this study, our main goal is to construct a modeling framework for the geometry of different dust components in SFGs using a statistical approach.  The Milky-Way like (MW-like) galaxies have stable and well-formed disks and the largest sample size in MGS, making them most suitable for this study. Therefore, we further select the SFGs with the stellar mass in the range $10^{10.2}-10^{10.6}M_\odot$. These selection criterics result our  final MW-like galaxy sample of 33,273 galaxies. We leave the exploration of the geometry of disk galaxies with other stellar masses to an upcoming study.

\subsection{Dust Reddening Measurements}

\subsubsection{Balmer decrement of emission lines \texorpdfstring{$E_{\rm{g}}$}{Eg} }
\label{sec:eg}

To isolate the effect of attenuation curves, we take the Balmer decrement, which compares the ratio of the intensities of  H$\alpha$ and  H$\beta$ emission lines $(f_{\rm{H}\alpha}/f_{\rm{H}\beta})_{\rm{obs}}$ with its intrinsic value $(f_{\rm{H}\alpha}/f_{\rm{H}\beta})_{\rm{int}}$ , to represent the nebular attenuation, 
    \begin{equation}
    E_{\rm{g}}(\rm{H}\alpha-\rm{H}\beta)=2.5\log\frac{(f_{\rm{H}\alpha}/f_{\rm{H}\beta})_{obs}}{(f_{\rm{H}\alpha}/f_{\rm{H}\beta})_{int}}\,,
    \label{taub}
    \end{equation}
where the intrinsic Balmer decrement $(f_{\rm{H}\alpha}/f_{\rm{H}\beta})_{int}$ is set to the typical Case B recombination value 2.86. Because the conversion from $E_{\rm{g}}(\rm{H}\alpha-\rm{H}\beta)$ to classical $E_{\rm{g}}(B-V)$ relies on the shape of the attenuation curve\footnote{For the classical Cazeltti law \citep{Calzetti1994}, $E_{\rm{g}}(B-V)= 0.9 E_{\rm{g}}(\rm{H}\alpha-\rm{H}\beta)$.}, we use $E_{\rm{g}}(\rm{H}\alpha-\rm{H}\beta)$ instead of classical $E_{\rm{g}}(B-V)$. For simplicity, we denote the nebular color excess $E_{\rm{g}}(\rm{H}\alpha-\rm{H}\beta)$ as $E_{\rm{g}}$ hereafter.  

We note that Equation \ref{taub} could result in negative color excess for some galaxies, which are likely to be caused by the measurement errors of the nebular emission line fluxes. For these galaxies, we set $E_{\rm{g}}=0$.

\subsubsection{Reddening of stellar population \texorpdfstring{$E_{\rm{s}}$}{Es}}
\label{sec:es}
    
We derive the dust attenuation of the stellar population by fitting the spectra of galaxies with SPS. In SPS, the observed spectrum of a galaxy is typically fit through
\begin{equation}
f(\lambda)=[\Sigma_{i=1,N}f_i * \rm{SSP}_i(\lambda)] \exp(-\tau_\lambda), 
\end{equation}
where the component in bracket is the sum of the fraction of each single stellar population (SSP), and $\tau_\lambda$ is the effective optical depth at wavelength $\lambda$ that parameterizes the global dust attenuation effect. Also, by assuming an attenuation curve for the stellar population, $k_{\rm{s}}(\lambda)$, the optical depth $\tau_\lambda$ is parameterized by
 \begin{equation}
\tau_{\lambda} = \frac{E_{\rm{s}}(B-V)}{1.086}\frac{k_{\rm{s}}(\lambda)}{k_{\rm{s}}(B)-k_{\rm{s}}(V)} \,
 \end{equation}
where $E_{\rm{s}}(B-V)$ is the global dust reddening of the stellar continuum to be fit. We use $E_{\rm{s}}(B-V)$ because it is less influenced by the attenuation curve shape than attenuation $A_{\rm{V}}$ (see more discussions in Section \ref{attenuationcurves}).

We use the full-spectrum stellar population fitting code STARLIGHT \citep[][]{Fernandes2005} to fit the stellar continuum (i.e. with emission lines masked) of each galaxy and derive its $E_{\rm{s}}(B-V)$ value. We adopt the BC03 stellar population \citep{Bruzual2003} for $\rm{SSP}_i(\lambda)$ and the standard Calzetti attenuation law \citep{Calzetti2000} for $k_{\rm{s}}(\lambda)$. Hereafter, we refer to the derived stellar color excess $E_{\rm{s}}(B-V)$ as $E_{\rm{s}}$.

There is a known dust-age-metallicity degeneracy effect in SPS fitting. As shown by \citet{Liniu2020}, among dust-age-metallicity, the dust attenuation feature is the one that can be most accurately and unbiased recovered from SPS fitting.  Moreover, since $E_{\rm{s}}(B-V)$ is a strong function of disk inclination, if its measurement were biased by the dust-age-metallicity degeneracy effect, we would expect  an inclination (dust attenuation) dependence of the average stellar population age and metallicity. We have checked this effect and find no inclination dependence (see appendix \ref{sec:z-m-a} for detail).

\subsection{\texorpdfstring{$E_{\rm{g}}$}{Eg} and \texorpdfstring{$E_{\rm{s}}$}{Es}}

\begin{figure*}[htbp]
\centering
\gridline{\fig{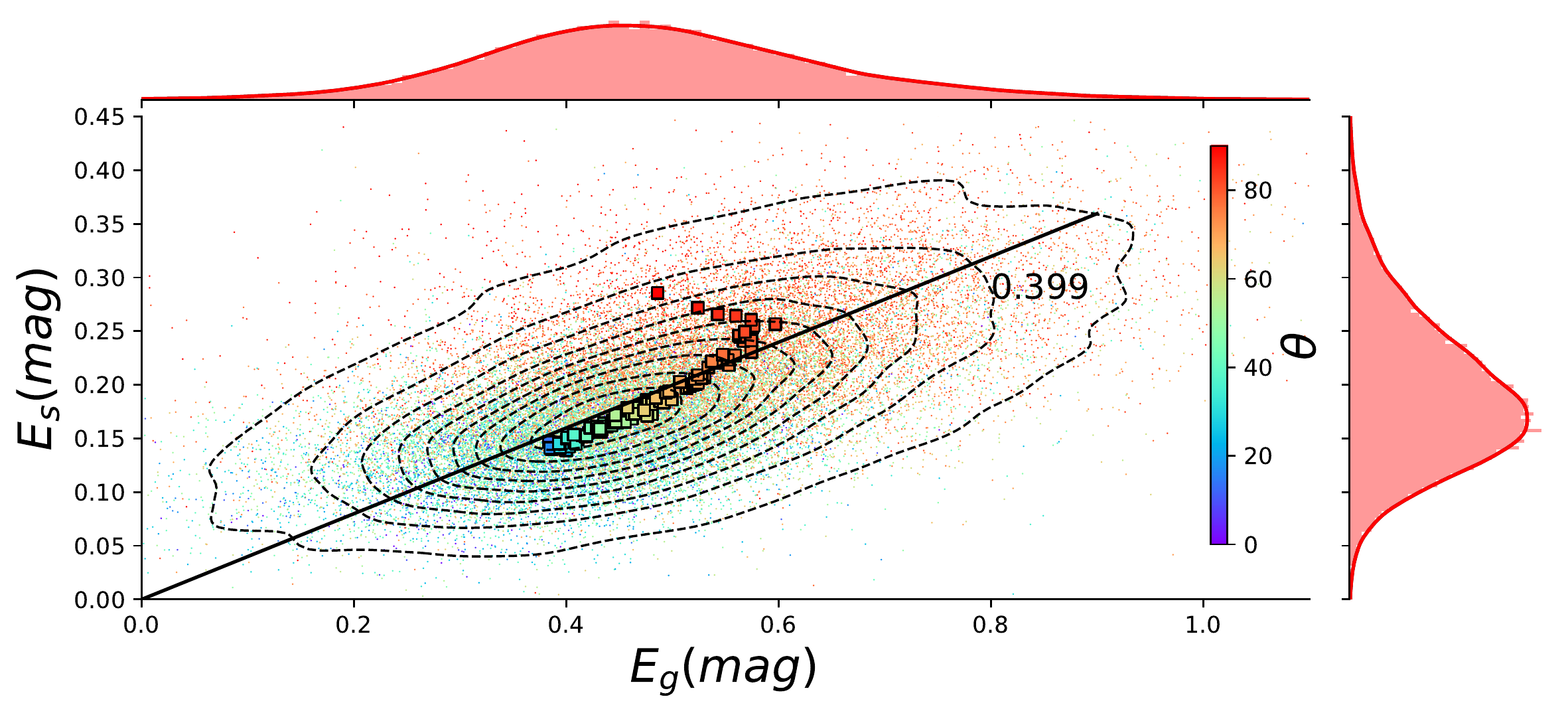}{6in}{(a) $E_{\rm{g}}-E_{\rm{s}}$ distribution}
          }
\gridline{\fig{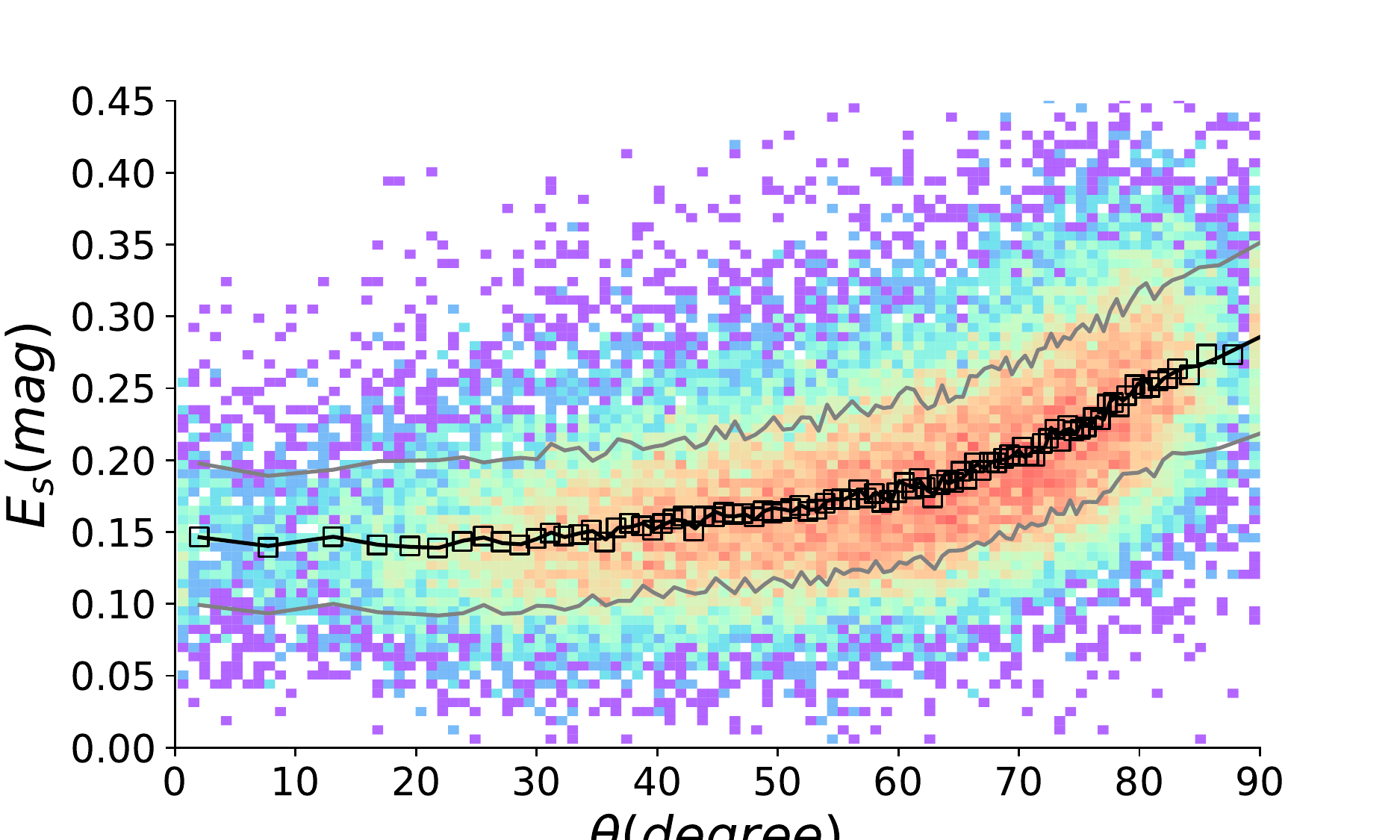}{3in}{(b) $E_{\rm{s}}-\theta$ distribution}
          \fig{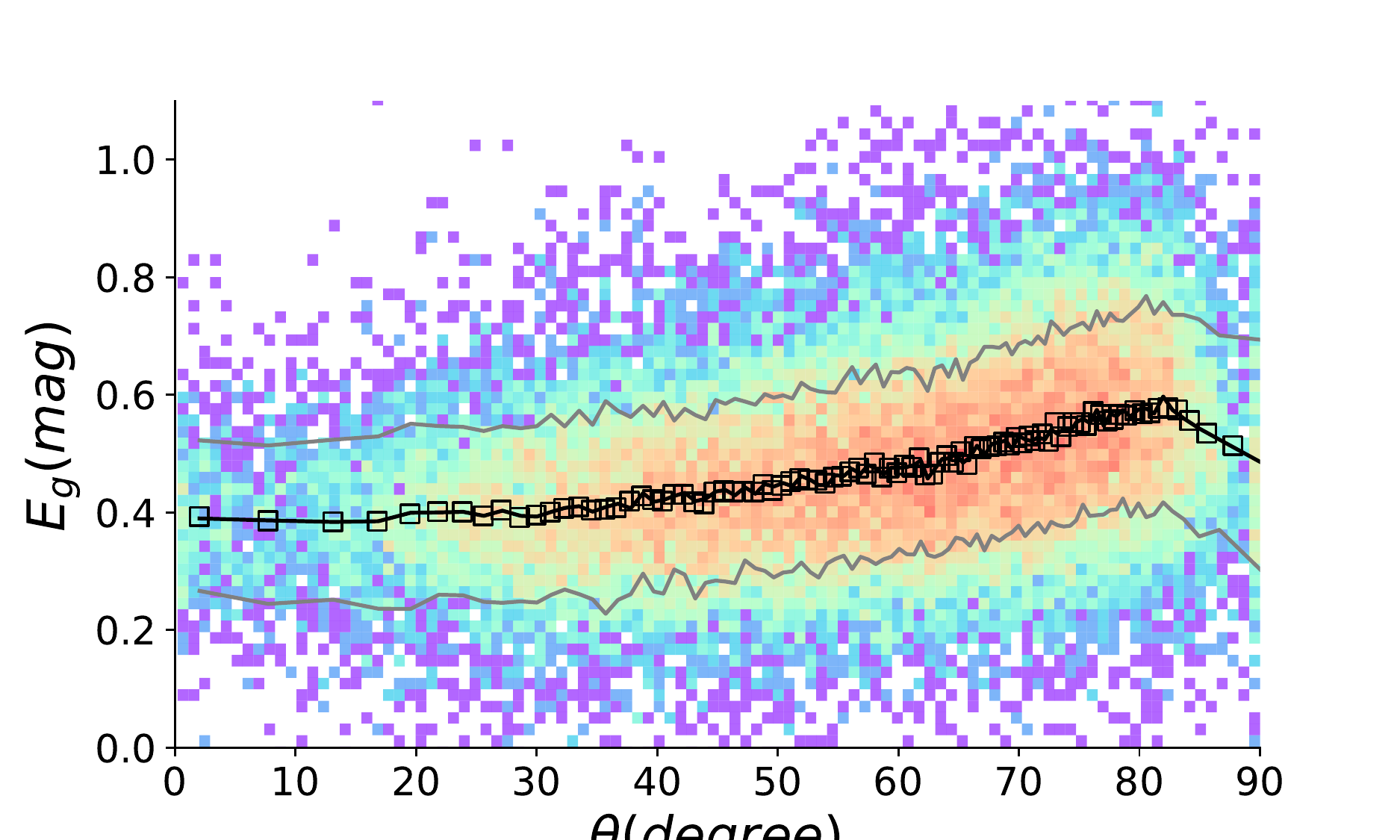}{3in}{(c) $E_{\rm{g}}-\theta$ distribution}
          }

	\caption{The distributions of  nebular color excess $E_{\rm{g}}$ and stellar color excess $E_{\rm{s}}$ as functions of disk inclination $\theta$. Panel (a): Joint distribution map of $E_{\rm{g}}$ and $E_{\rm{s}}$ of our sample galaxies, where $\theta$ is indicated by the color. The black line shows the regression line with  forced zero intercept, $E_{\rm{g}}=0.4E_{\rm{s}}$. Panel (b):  $E_{\rm{s}}$ as function $\theta$, where the median of $E_{\rm{s}}$ distribution at each $\theta$ bin are shown as  square dots connected with a solid line and the 16 and 84 percentiles are indicated by two solid curves. Panel (c):  $E_{\rm{g}}$ as function $\theta$,  the same structure as Panel (b).}
	\label{egest}
   \end{figure*}
    
We show the joint distribution of $E_{\rm{g}}$ and $E_{\rm{s}}$ for our sample galaxies in the top panel of Figure \ref{egest}, where we have removed 23 outliers with $E_{\rm{g}}$ or $E_{\rm{s}}$ values deviate more than 5$\sigma$ from their median values. The final sample contains 33,250 galaxies. As can be seen, $E_{\rm{g}}$ spans a wide range from 0 to $\sim 1$ mag with a median value of $\sim0.45$ mag, while $E_{\rm{s}}$ is mostly distributed in the range 0-0.4 mag and with a median value 0.17 mag.

In general, $E_{\rm{g}}$ and $E_{\rm{s}}$  show a monotonic correlation and with a mean ratio of $E_{\rm{s}}/E_{\rm{g}} \sim 0.4$. Considering the conversion factor from $E_{\rm{g}}$ to $E_{\rm{g}}(B-V)$ ($\sim 0.9$ for the Calzetti attenuation curve),  this mean ratio is consistent with the conical value $f\sim 0.44$   of  \citet{Calzetti2000}.
 
\subsection{Inclination Dependence of Dust Attenuation}

We show the inclination dependence of $E_{\rm{s}}$ and $E_{\rm{g}}$  for our MW-like galaxies by color-coding each galaxy with their disk inclination $\theta$ in the top panel of  Figure \ref{egest}. As expected,  there is a general trend that the heavily attenuated galaxies are mostly these galaxies with large inclination angle $\theta$. However, at given $\theta$, there are  large scatters on both $E_{\rm{g}}$ and $E_{\rm{s}}$, which reflects the variation of the intrinsic dust attenuation among different galaxies.

To better characterize the inclination dependence of these two types of dust reddening, we divide the sample galaxies into 90 $\theta$ bins and require that  each $\theta$ bin includes the same number of galaxies. We show the  median values (square dots connected with a solid line)  of $E_{\rm{s}}$ and $E_{\rm{g}}$ for each $\theta$  bin in the bottom left and bottom right panels of Figure \ref{egest}, where the $16$ and $84$ percentiles of  the $E_{\rm{g}}$ and $E_{\rm{s}}$ distribution are indicated by the gray solid lines. Because of the large number of the galaxies in each $\theta$ bin ($n\sim 300$), the uncertainty of the median values of $E_{\rm{g}}$ and $E_{\rm{s}}$ are both smaller than 0.01.

As shown in  Figure \ref{egest}, at inclination $\theta < 75^\circ$, both  $E_{\rm{g}}$ and $E_{\rm{s}}$ increase monotonically with the disk inclination $\theta$, while $E_{\rm{g}}$ shows a sharper increase trend. For highly inclined disk galaxies ($\theta>75^\circ$), the trend of $E_{\rm{s}}$ becomes flat, whereas the trend of $E_{\rm{g}}$ is reversed, i.e. $E_{\rm{g}}$ decreases with increasing $\theta$. 

The monotonic increase trend of $E_{\rm{s}}$ as a function of $\theta$ has been reported by many studies, and can be reasonably explained by either simple parametric models or dedicated RT models \citep[e.g.][]{Shao2007, Maller2009, Masters2010, Yip2010}. Different trends of $E_{\rm{g}}$ and $E_{\rm{s}}$ as functions of disk inclination also have been found in observations \citep[e.g.][]{Yip2010, Chevallard2013, Battisti2017}. In our study, because of the equal-size binning  of $\theta$ and the large sample size, for the first time, we reveal a complicate inclination dependence of $E_{\rm{g}}$, especially at high $\theta$ values.

\section{Modelling of inclination dependence of \texorpdfstring{$E_{\rm{g}}$}{Eg} and \texorpdfstring{$E_{\rm{s}}$}{Es}: simple models}
\label{simple model}
From this section, we aim to build geometric models for disk galaxies so as to model the observed inclination dependence of $E_{\rm{s}}$ and $E_{\rm{g}}$ shown in the bottom-left and bottom-right panels of Figure \ref{egest} . We start with the two most frequently used simple models, the uniform mixture model and the screen model, to model the observed $E_{\rm{s}}$ (Section \ref{e_s mix}) and $E_{\rm{g}}$ (Section \ref{e_g screen}), respectively. Before that, we introduce the common parts of the two different dust attenuation models.

We assume that our MW-like galaxies have the same geometry configuration so that the inclination dependence of attenuation being observed is only caused by the angle of view. In our modelling, we assume that the main components of disk galaxies follow a double exponential model in geometry,
\begin{equation}
D_{\rm{comp}}(r,h)=\exp(-\frac{r}{R_{\rm{comp}}}-\frac{\vert h \vert}{h_{\rm{comp}}}),
\label{Dexp}
\end{equation}
where $R_{\rm{comp}}$, $h_{\rm{comp}}$ are exponential disk scale-length and scale-height, respectively.  Here, depending on where the equation is used, the component could  either be referred to emission source (e.g. stars) or dust component. Moreover, since the SDSS fiber spectra only cover the central part of the observed galaxies, we also consider only the attenuation of the line of sight through the center of the model galaxy. For this line of sight, the integration of a double exponential component has inclination dependence 

\begin{equation}
F_{\rm{comp}}(\theta) \propto \frac{h_{\rm{comp}}/R_{\rm{comp}}}{\cos\theta+h_{\rm{comp}}/R_{\rm{comp}}\sin\theta} \,.
\label{Fcomp}
\end{equation}
For convenience, we set $F_{\rm{comp}}=1$ when galaxies are viewed as edge-on ($\theta=90^\circ$).

In addition, we make a convention for the subscripts of the symbols to be used. The order of the subscripts is `wavelength', `component' and `specific'. The wavelength term is typically written as `$\lambda$' . If  not explicitly specified,  the default wavelength is at V-band.  The `component' term includes `stellar disk',`nebular disk', and `clumps', which are denoted by`s', `g ', `cl' respectively. The `specific'  term includes`central', `total', `un-attenuated', denoted by `0', `A', and `*', respectively. 

\begin{figure}[htbp]
\centering
\gridline{\fig{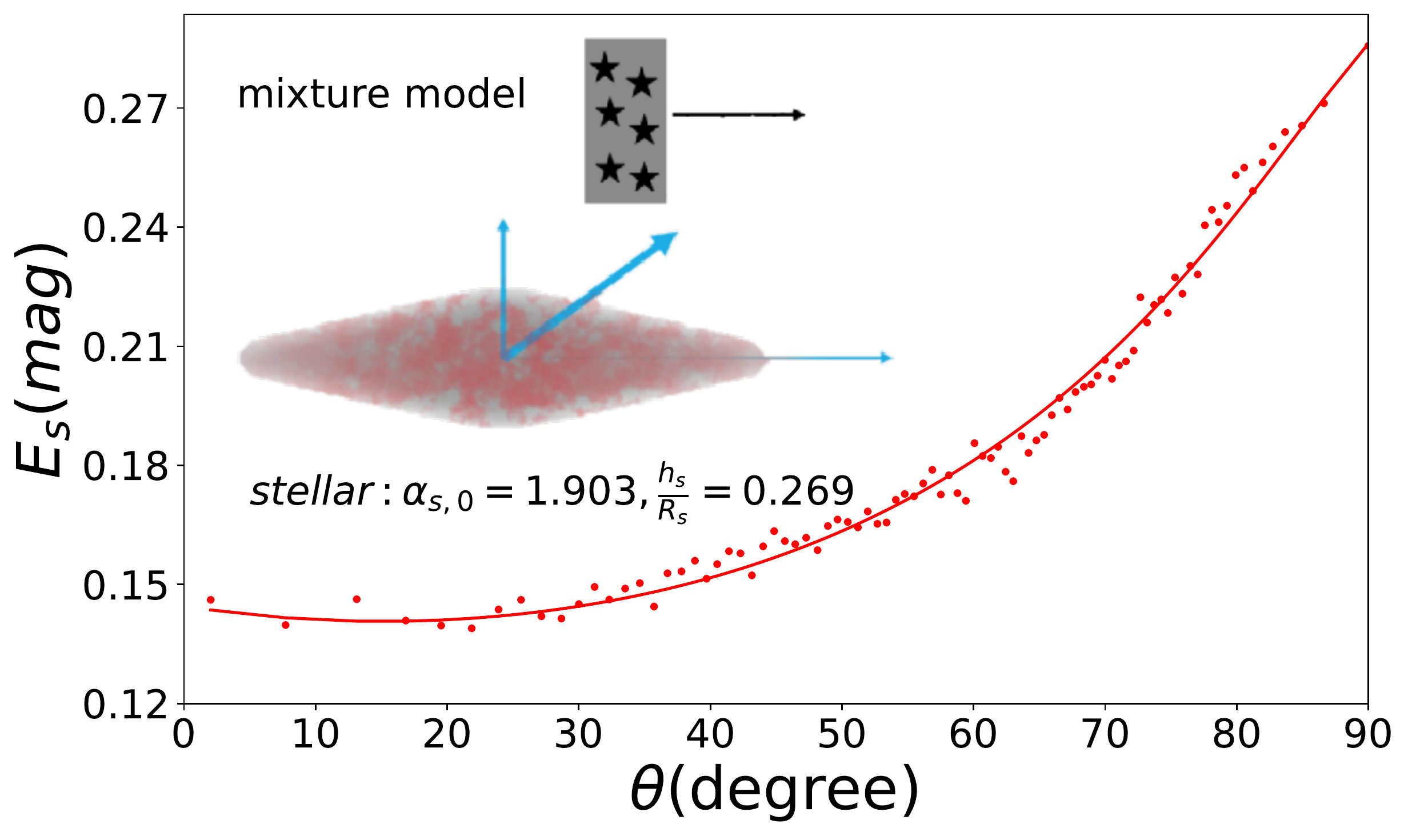}{3in}{(a) mixture model}
          }

\gridline{\fig{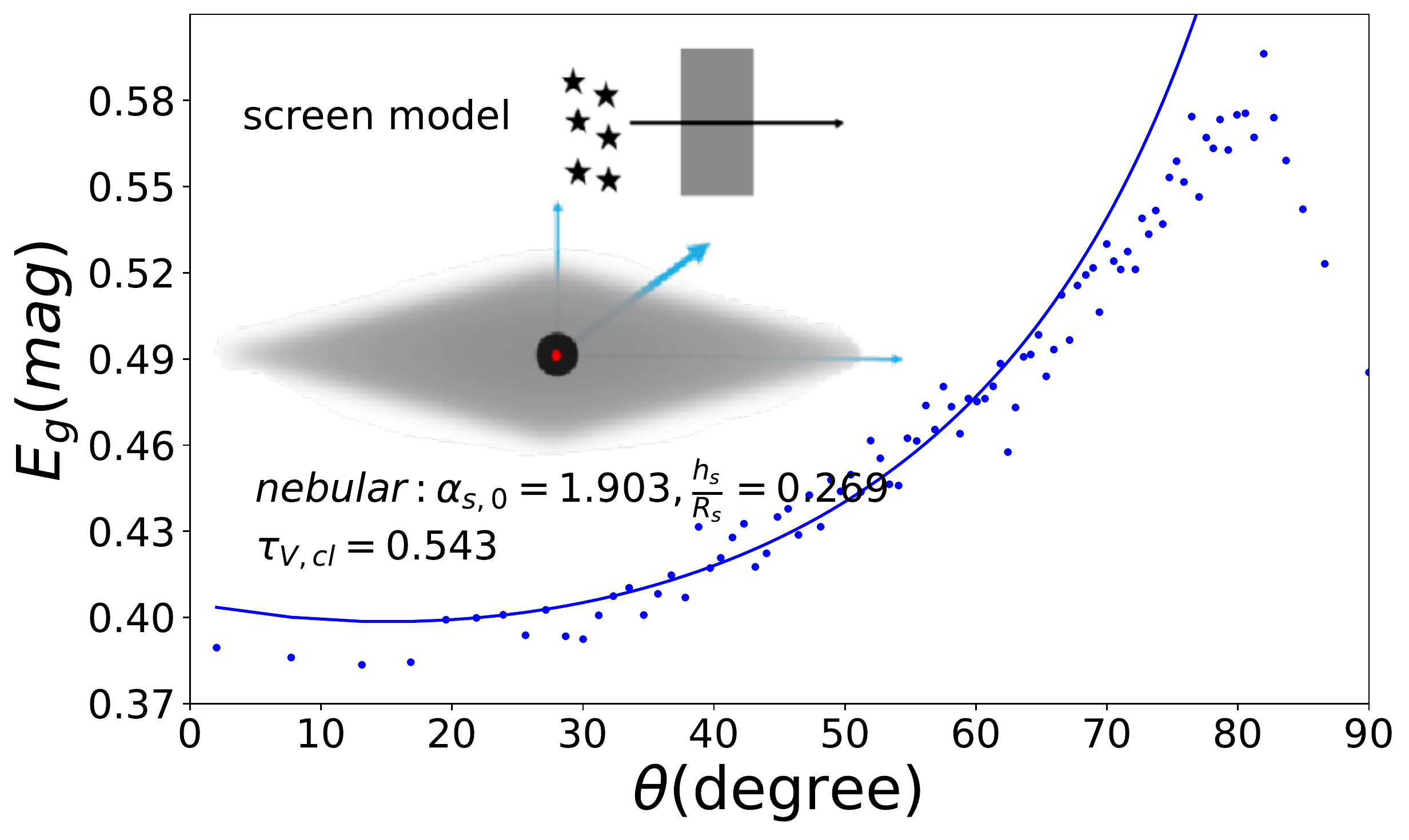}{3in}{(b) screen model}
          }

    \caption{Modelling of the inclination dependence of $E_{\rm{s}}$ and $E_{\rm{g}}$ with simple models. Panel (a): uniform mixture model for $E_{\rm{s}}$. The small dots show the observed median $E_{\rm{s}}-\theta$ relation, while the solid line shows the best model fitting presented in Section \ref{e_s mix}. 
    Panel (b): screen model for $E_{\rm{g}}$. The small dots show the observed median $E_{\rm{g}}-\theta$ relation, while the solid line shows the best model fitting presented in Section \ref{e_g screen}. The schematic diagram of each model and the values of the best model parameters are inserted in each panel.}
    \label{simple_model}
 \end{figure}

\subsection{uniform mixture Model for \texorpdfstring{$E_{\rm{s}}$}{Es}}
\label{e_s mix}

On galactic scale, the dust attenuation of diffuse ISM dust on stellar emission can be simplified and parameterized by a uniform mixture model. We show  a schematic map for such a simple mixture model in the top panel of  Figure \ref{simple_model}.  In specific of a disk galaxy, we assume that the density profiles of stellar population and  diffuse ISM dust  follow the same exponential disk model (Equation \ref{Dexp}),
\begin{equation}
 \rho_{\rm{s}}(r,h)=\rho_{\rm{s,0}}D_{\rm{s}}(r,h|R_{\rm{s}},h_{\rm{s}}), 
 \label{rho_star}
\end{equation}
where  $R_{\rm{s}}$ and $h_{\rm{s}}$ are the scale-length and scale-height of the stellar disk, $\rho_{\rm{s,0}}$ is the central effective density of dust particles or stellar population, depending on the usage of this Equation. 

With this exponential disk modelling, for a disk galaxy  with inclination $\theta$, the total line of sight optical depth along the galactic center direction  is 

\begin{equation}
	\tau_{\lambda,\rm{s,A}}(\theta)=\int \rho_{\rm{s}}\kappa_{\lambda} \rm{d}s = 2 \rho_{\rm{s,0}}\kappa_{\lambda}R_{\rm{s}}F_{\rm{s}}(\theta),
    \label{exp_tau0}
\end{equation}
where $\kappa_\lambda$ is the dust extinction coefficient (absorption cross-section) at given wavelength $\lambda$, and $F_{\rm{s}}(\theta)$ is the inclination dependence term shown in Equation \ref{Fcomp}.  We parameterize the wavelength dependence of dust extinction coefficient, namely the dust extinction curve $\kappa(\lambda)$ with a simple power-law,

\begin{equation}
\kappa(\lambda)=\kappa_{V}(\frac{\lambda}{5500\text{\AA}})^{-\beta}\,
\label{power curve}
\end{equation}
which is normalized by $V$ band extinction coefficient  $\kappa_V$ at  its effective wavelength $5500\text{\AA}$. We set the power-law index $\beta$  to be $1.32$, so that it has $R_V= \frac{A_V}{E(B-V)}=3.1$, a typical value for the diffuse ISM dust in the Milky-Way \citep{Weingartner2001, Fitzpatrick2019, Li2017}.  

Moreover, during the calculation of $\tau_{\lambda,\rm{s,A}}(\theta)$, $\kappa_{V}$  is coupled with the central dust density parameter $\rho_{\rm{s,0}}$. Therefore, for simplicity, we define a new parameter $\alpha_{\rm{s,0}} \equiv \rho_{\rm{s,0}} \kappa_{V}$, which directly represents the central dust absorption density. With this new parameter, we have

\begin{equation}
	\tau_{\lambda,\rm{s,A}}(\theta)= 2 \alpha_{\rm{s,0}}(\frac{\lambda}{5500\text{\AA}})^{-1.32} R_{\rm{s}}F_{\rm{s}}(\theta).
    \label{exp_tau}
\end{equation}
For the uniform mixture model,  the total emitted intensity $I$ is linked to the total unextincted intensity $I_*$ through 
\begin{equation}
I=I_*\frac{1-e^{-\tau_{\lambda,\rm{s,A}}}}{\tau_{\lambda,\rm{s,A}}},
\label{mix_tau}
\end{equation}
where  $\tau_{\lambda,\rm{s,A}}$ is the total optical depth along the given line of sight.  The dust reddening then follows

\begin{equation}
E_{\rm{s}} (\lambda_1-\lambda_2)=-2.5\log(\frac{I_{\lambda_1}/I_{\lambda_2}}{I_{\lambda_1,*}/I_{\lambda_2,*}})=2.5\log(\frac{\tau_{\lambda_1}}{\tau_{\lambda_2}}\frac{1-e^{-\tau_{\lambda_2}}}{1-e^{-\tau_{\lambda_1}}})  \,.
\label{mix_e}
\end{equation}
As shown by Equations \ref{mix_tau} and \ref{mix_e},  for the uniform mixture model, when the optical depth is thick ($\tau>> 1$), there is a simple correlation of emitted intensity with $\tau$  ($I\approx I_*/\tau$) and a nearly constant color excess $E(\lambda_1-\lambda_2)\approx $ $2.5\rm{log}(\tau_{\lambda_1}/\tau_{\lambda_2})$. 
      
The optical depth $\tau$ of the uniform mixture model (Equation \ref{exp_tau}) is composed of two terms. One is that describing the global amount of dust attenuation, which could be parameterized by the $V$ band dust optical depth to the galactic center in edge-on case, $\tau_{\rm{s,A}}(90^\circ)\equiv 2\alpha_{\rm{s,0}}R_{\rm{s}}$.  The other is the inclination dependence term $F_{\rm{s}}(\theta)$  (Equation \ref{Fcomp}), which has the only parameter $h_{\rm{s}}/R_{\rm{s}}$, i.e., the disk scale-height to scale-length ratio. Since the disk scale-length $R_{\rm{s}}$ is included in both terms, for simplicity, we set $R_{\rm{s}}=1$. Then, the two fitting model parameters become $\alpha_{\rm{s,0}}$ and $h_{\rm{s}}$ , which are in  units of $R_{\rm{s}}^{-1}$ and $R_{\rm{s}}$, respectively. We fit the observed inclination dependence of $E_{\rm{s}}$ with these two free model parameters. The best fitting of the $E_{\rm{s}}-\theta$ relation is shown as the red solid line in the top panel of  Figure \ref{simple_model} and the two best model parameters are listed in Table \ref{2com}.

Figure \ref{simple_model} (top panel) shows that this uniform mixture model fits the observed $E_{\rm{s}}-\theta$ relation quite well. In our best fitting, the disk component has a height-to-radius ratio $h_{\rm{s}}/R_{\rm{s}}= 0.27$. The best estimate of $\alpha_{\rm{s,0}}$  is $1.9$, which means that the global $V$ band optical depth along the galaxy center direction is varying from $\sim 1$ (face-on) to $\sim 4$ (edge-on), corresponding to the  effective dust attenuation from $A_{V}\sim 0.50 $ mag (face-on) to $1.41$ mag (edge-on), which is broadly consistent with the observational values  \citep[$0.5-1.5$ mag,][]{Bianchi2007,DeGeyter2014,Casasola2017}.

\subsection{Screen  Model for \texorpdfstring{$E_{\rm{g}}$}{Eg} }
\label{e_g screen}

 The nebular emission of star-forming (HII) regions is known to be distributed on a much thinner disk than the stellar component \citep[e.g.][]{Anderson2019}.  Considering that the SDSS fiber spectra only target the central regions of galaxies,  these HII regions might be viewed as a point-like source  in galaxy center if they do not overlap each other along the line of sight.  In this simplified model, the nebular emission lines emitted from the central star-forming regions are then only subject to the extinction of outer dust layers along the line of sight, which is  consistent with a screen model. 

For the dust screen model,  the observed intensity of an object with intrinsic intensity $I_0$ is determined by the optical depth $\tau$ of the dust layer along the line of sight, $I=I_0e^{-\tau}$.  We show a schematic map for this simple  dust screen model in the bottom panel of  Figure \ref{simple_model}.  The line of sight dust consists of two parts. One is the dust component in diffuse ISM and the other is the dust shell of nebular region itself. For the former component, we take the best model estimate of Section \ref{e_s mix}, i.e. an exponential diffuse ISM dust disk layer with geometric configuration $h_{\rm{s}}/R_{\rm{s}}=0.27$ and the central dust absorption density $\alpha_{\rm{s,0}}=1.9$.  Since the nebular emission region is assumed to be located at  galaxy center, the inclination dependence of its line of sight optical depth from diffuse ISM dust is therefore half of the $\tau_{\lambda, s,A}(\theta)$ that parameterized by Equation \ref{exp_tau}. For the dust component of the nebular emission region itself, we assume that the dust layer is distributed in a spherical shell. In this case, the dust optical depth of the shell has no inclination dependence and therefore can be parameterized by a  constant $\tau_{\lambda,\rm{cl}}$. Putting these two dust components together,  for disk galaxies with  inclination $\theta$, the total line of sight optical depth of the modelled nebular emission line region in galaxy center is
 \begin{equation}
	\tau_{\lambda,\rm{g,A}}(\theta)=\frac{\tau_{\lambda,\rm{s,A}}(\theta)}{2}+\tau_{\lambda,\rm{cl}}  \,.
    \label{exp+sph}
\end{equation}
We assume that the extinction curve of the dust particles in nebular regions is the same as that of diffuse ISM dust (Equation \ref{power curve}) and then parameterize  $\tau_{\lambda,\rm{cl}}$  with the $V$ band optical depth $\tau_{\rm{cl}}$. In this simple dust screen model,  the observed color excess of emission lines is 
\begin{equation}
E_{\rm{g}}(\lambda_1-\lambda_2)=-2.5\log(\frac{I_{\lambda_1}/I_{\lambda_2}}{I_{\lambda_1,*}/I_{\lambda_2,*}})=1.086(\tau_{\lambda_1,g,A}-\tau_{\lambda_2,g,A}) \,.
\label{screen eq}
\end{equation}    
With above  model, we fit the observed inclination dependence of $E_{\rm{g}}$ shown in the bottom right panel of Figure \ref{egest} with the only parameter $\tau_{\rm{cl}}$. We obtain the best fit with $\tau_{\rm{cl}}=0.54$, which is shown as the solid line in the bottom panel of Figure \ref{simple_model} and listed in Table \ref{2com}. As can be seen, by including the dust component in nebular regions, the dust screen model of Equation \ref{exp+sph} generally reproduces the observed $E_{\rm{g}}-\theta$ relation for  low inclination disks ($\theta < 70^\circ$). However, for these high inclined disks ($\theta > 70^\circ$), the observed  $E_{\rm{g}}-\theta$ relation shows a saturation effect, which  can not be accounted by our simple dust screen model. This saturation effect resembles the behavior of a mixture model in optically thick case. That means, at very high inclinations, these nebular emission regions along the line of sight can no more be approximated by a point source in galaxy center, which suggests a more complex geometric configuration of these nebular  line regions.

\section{Modelling of inclination dependence of \texorpdfstring{$E_{\rm{g}}$}{Eg} and \texorpdfstring{$E_{\rm{s}}$}{Es} simultaneously: \textit{Chocolate Chip Cookie }Model}
\label{CCC model}

\begin{figure*}[htb]

    \centering
\gridline{\fig{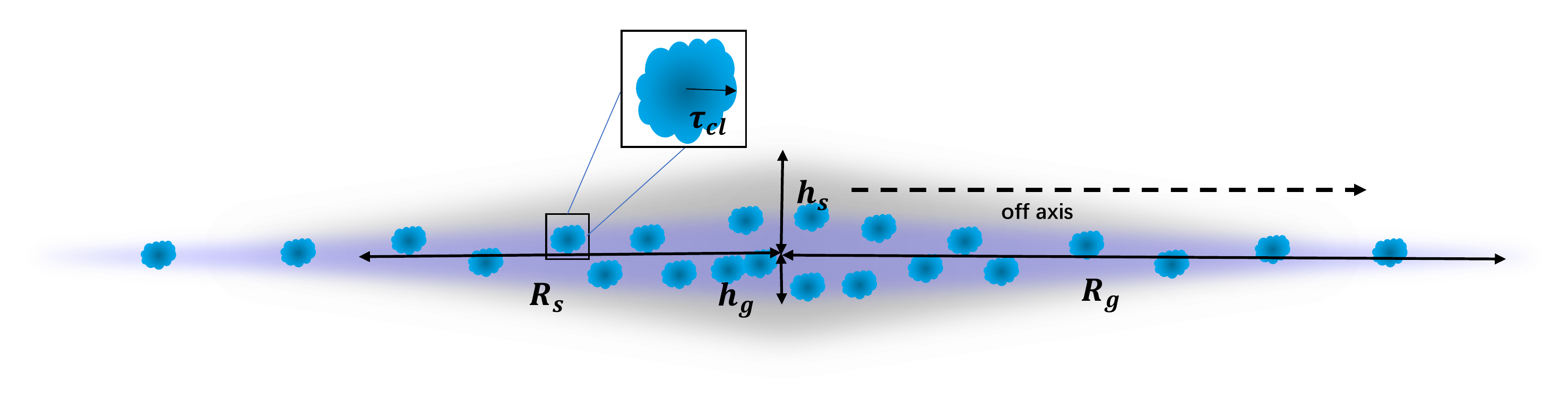}{5in}{(a) "\textit{Chocolate Chip Cookie}" model} 
          }
\gridline{\fig{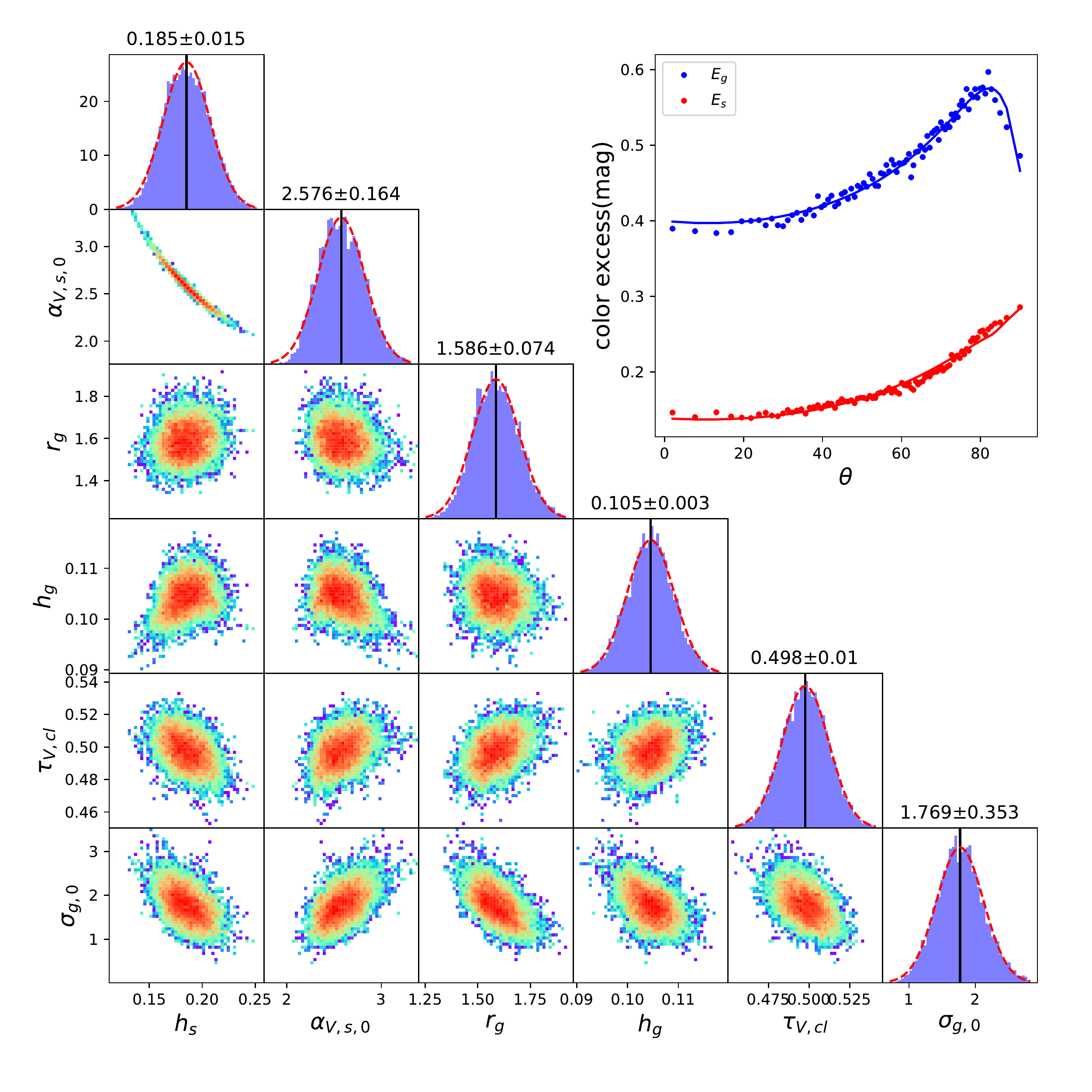}{5in}{(b) MCMC fitting results}}

    \caption{Modeling of the $E_{\rm{g}}-\theta$ and $E_{\rm{s}}-\theta$ relations with the \textit{ Chocolate Chip Cookie} (CCC) model.  Panel (a) : schematic map of the CCC model, where $R_{\rm{g}},h_{\rm{g}},R_{\rm{s}},h_{\rm{s}}$ are the scale-length and scale-height of the clumpy nebular disk and diffuse ISM disk respectively and $\tau_{\rm{cl}}$ is the optical depth of each individual clumps. The dashed arrow line shows the off-axis effect of a line of sight, where the stellar emission has not been extincted by the nebular disk (see Section \ref{Off-axis effect} for detail). Panel (b): the corner plot of the MCMC fitting for the six free model parameters (Table \ref{2com}) of the CCC model, where the best fittings of  $E_{\rm{g}}-\theta$ and $E_{\rm{s}}-\theta$ relations are inserted at the top right corner of this panel.} 
    \label{CCC_model}
 \end{figure*}

\begin{table*}[htbp]
\caption{The parameters of mixture model (Section \ref{e_s mix}), screen model (Section \ref{e_g screen}), and  \textit{Chocolate Chip Cookie} model (Section \ref{CCC model}).}
\label{2com}
\centering
\begin{threeparttable}
	\begin{tabular}{cccccc}
	\hline\hline
model& component&parameter&description& scaled value&physical value\\
\hline\hline
\multirow{2}{*}{\makecell{mixture \\ model}} & \multirow{2}{*}{stellar} &$ h_{\rm{s}}$ &  scale-height of stellar disk&0.27 $R_{\rm{s}}$ & - \\
\cline{3-6}
&& $\alpha_{\rm{s,0}}$ &central dust absorption density&1.9 $R_{\rm{s}}^{-1}$&- \\
\hline
\multirow{3}{*}{\makecell{screen \\ model}}    & \multirow{2}{*}{stellar} & $h_{\rm{s}}$&scale-height of stellar disk&0.27 $R_{\rm{s}}$ & - \\
\cline{3-6}
&& $\alpha_{\rm{s,0}}$  &central dust absorption density in $V$ band&1.9 $R_{\rm{s}}^{-1}$&-\\
 \cline{2-6}
& HII region & $\tau_{\rm{cl}}$ & centric $V$ band optical depth of HII region& 0.54  & -\\
\hline  
\multirow{11}{*}{\makecell{\textit{Chocolate} \\ \textit{Chip} \\ \textit{Cookie} \\
model }}    &   \multirow{4}*{\makecell{stellar \\ disk}}& $h_{\rm{s}}$&scale-height of stellar disk& 0.19 $R_{\rm{s}}$&0.41 kpc\\
\cline{3-6}
 & &$R_{\rm{s}}$&scale-length of stellar disk&set as unit size&2.1 kpc\\
\cline{3-6}
& &$\alpha_{\rm{s,0}}$& stellar disk $V$ band central absorption density&2.58 $R_{\rm{s}}^{-1}$&1.22 kpc$^{-1}$\\
\cline{3-6}
&   &$I_{\lambda,s}$&emissivity of stellar disk & set as 1 & - \\
\cline{2-6}
&     \multirow{7}*{\makecell{clumpy \\ nebular \\ disk}}&$h_{\rm{g}}$&scale-height of clumpy nebular disk& 0.11 $R_{\rm{s}}$&0.22 kpc\\
\cline{3-6}
&    &$R_{\rm{g}}$&scale-length of clumpy nebular disk&1.59 $R_{\rm{s}} $ & 3.33 kpc\\
\cline{3-6}
&  & $\rho_{\rm{g,0}}$ & central number density of clumps& -\tnote{*}  & $\sim 2,700$ kpc$^{-3}$ \\
\cline{3-6}
&   & $I_{\lambda,g}$ & emissivity of nebular disk & set as 1& -\\
\cline{3-6}
&   & $\tau_{\rm{cl}}$ & centric $V$ band optical depth of clumps& 0.50 & 0.50\\
\cline{3-6}
&  & $R_{\rm{cl}}$ & effective radius of a  clump&-\tnote{*} & $\sim 30$ pc\\
\cline{3-6}
& & $\sigma_{\rm{g,0}}$& central cross-section of clumps& 1.77 $R_{\rm{s}}^{-1}$&0.84 kpc$^{-1}$\\
\hline\hline
	\end{tabular}
     \begin{tablenotes}
 \footnotesize
       \item[*] $\rho_{\rm{g,0}}$  and $R_{\rm{cl}}$ are reduced as $\sigma_{\rm{g,0}}$ ($\sigma_{\rm{g,0}}=\rho_{\rm{g,0}}\pi R^2_{\rm{cl}}$).
     \end{tablenotes}
     \end{threeparttable}
   \end{table*}

As shown above in Section \ref{e_g screen}, a simple diffuse ISM dust component plus a simple spherical HII region can not properly account the observed  $E_{\rm{g}}-\theta$ relation, especially  for highly inclined disks. Moreover, the nebular dust component in principle will also have extinction effect on the stellar continua, which has not been taken into account by our simple mixture model presented in Section \ref{e_s mix}. In this section, we present a self-consistent two dust component model and use it to  model the observed inclination dependence of the two different dust reddening features simultaneously.

As  shown in Section \ref{e_s mix}, a uniform mixture model could generally reproduces the inclination dependence of the stellar reddening effect.  Therefore, we still assume that the  stellar emission comes mainly from a  \textbf{stellar disk} , while the diffuse ISM dust component is uniformly mixed in. Indeed, there are observations show diffuse far infrared (FIR) emissions at high latitude of edge-on disk galaxies, whose scale-length and scale-height are linearly correlated with that of the stellar component \citep[][and references therein]{Mosenkov2022}.

The nebular emissions of SFGs are known to be mainly contributed by clumpy distributed HII regions. Because of heavy self-attenuation and short life-scales, the HII regions of local SFGs contribute few percent of stellar emissions. Observations show that the global structure of these HII regions (the youngest stellar population) also follows a disk geometry, but thinner and more extended than the general stellar populations \citep[e.g.][]{Bobylev2021, Monteiro2021, Anderson2019}. To approximate the geometry configuration of these clumpy HII regions, we assume that they are spherically symmetric and physically identical.  The number density of these clumpy regions then follows another exponential disk distribution, which is co-planar and concentric with the \textbf{stellar disk}. We refer this geometry configuration of the clumpy HII regions as the \textbf{clumpy nebular disk} hereafter. 

Now, we have two different disk components for our model galaxy. One is the \textbf{stellar disk} composed of continuous stellar emission and its associated diffuse ISM dust, which has already been defined by Equation \ref{rho_star} in Section \ref{e_s mix}; the other is the  \textbf{clumpy nebular disk}, composed of HII regions, each of which has an intrinsic optical depth $\tau_{\rm{cl}}$ for its nebular emission $I_{\rm{g}}$ additionally. We refer to this two dust component model as the \textit{Chocolate Chip Cookie} model (hereafter CCC model), where the dusty clumpy HII regions (“chocolate chip”) are embedded in a stellar disk (“cookie”).  The schema of  our newly proposed CCC model is shown in the top panel of Figure \ref{CCC_model}.

More specifically, we also parameterize the \textbf{clumpy nebular disk}  as an exponential disk,
\begin{equation}
\rho_{\rm{g}}(r,h)=\rho_{\rm{g,0}}D_{\rm{g}}(r,h|R_{\rm{g}},h_{\rm{g}}),
\label{rho_gas}
\end{equation}
where $\rho_{\rm{g,0}}$  now is the central number density of clumpy regions, $R_{\rm{g}},h_{\rm{g}}$  represent the nebular disk scale-length and scale-height respectively.   However, unlike the continuous distribution of \textbf{stellar disk}, the distribution of  \textbf{nebular regions} is clumpy, and the line of sight optical depth of  clumpy regions  is not a simple integration  of Equation \ref{rho_gas}. Moreover, these clumpy distributed regions would also have dust extinction effects on the stellar emission. Therefore, we need further simplifications to model the dust attenuation effect of the clumpy regions on both of the nebular emission and stellar emission, which we will discuss in the following subsections.

\subsection{Modelling of the Clumpy Regions}
\label{Modelling of the clumpy regions}

We assume that  each clumpy region is self-extincted by its dust shell with optical depth $\tau_{\rm{cl}}$. When a clumpy region is in front of a emission source (e.g. a star or another clumpy region) along the line of sight, its mean (effective) optical depth is
\begin{equation}
\bar{\tau}_{\rm{cl}}=\frac{\int_0^{R_{\rm{cl}}} 4\pi r \sqrt{R_{\rm{cl}}^2-r^2}\tau_{\rm{cl}} \rm{d}r}{\pi R_{\rm{cl}}^2}=\frac{4}{3}\tau_{\rm{cl}} \,,
\label{eff_tau}
\end{equation}
where $R_{\rm{cl}}$ is the radius of a single clumpy region. We note that the optical depth $\tau$ is always a function of wavelength $\lambda$ (Equation \ref{power curve}). Therefore, we henceforth omit the subscript $\lambda$ for $\tau_{\lambda}$ to avoid complexity of notation.

In our sample,  the flux of a galaxy is collected from the SDSS fiber aperture with a radius of 1.5 arcsec\footnote{For SDSS I and II, the fibers of SDSS spectrograph have a diameter of 3 arcsec, while for SDSS III $\&$ IV, the BOSS spectrograph has a fiber diameter of 2 arcsec.}, which corresponds to a radius of $ R_{\rm{a}}\sim 2.2$ kpc at redshift $z \sim 0.07$ (the median redshift of our MW-like galaxy sample). We parameterize the number of clumpy regions in each SDSS aperture as $N \propto n_{\rm{g}}{\pi}R_{\rm{a}}^2$ , where $n_{\rm{g}}$ is the column number density of clumpy regions that equals to the integrate of $\rho_{\rm{g}}$ along the line of sight. We note  $R_{\rm{cl}}$ is far smaller than $R_{\rm{a}}$ and define a covering factor $p$ as the ratio of the cross-sectional area of each clump to the aperture area, 
\begin{equation}
p \sim \frac{R_{\rm{cl}}^2}{R_{\rm{a}}^2}\,,
\end{equation}
which also indicates the probability of a random emission source being covered by a foreground clumpy region along the line of sight.  Because $R_{\rm{cl}}<<R_{\rm{a}}$, the covering factor $p<<1$.

We assume that there are $N$ discrete clumpy regions  randomly distributed within the aperture. Then, the number of foreground clumpy regions along a particular line of sight obeys a binomial distribution,

\begin{equation}
B(k|N,p)=\frac{N!}{k!(N-k)!}p^k p^{N-k} \,.
\end{equation}
Therefore, for an emitting source with  intrinsic flux density $I_e$, the observed flux density after extinction by these front clumpy regions is  

\begin{equation}
I_\mathrm{ext,cl}=\sum^N_{k=0} I_eB(k|N,p)*\exp(-k\bar{\tau}_{\rm{cl}}) \,.
\label{fluxBio}
\end{equation}

When $\bar{\tau}_{\rm{cl}}<1$ and $p<<1$, the discrete binomial distribution can be approximated by a continuous Gaussian distribution ($\mathcal{N}$) with mean $\mu=Np$ and variance $\sigma^2=Np(1-p)$ . Then, we obtain an approximation 
\begin{equation}
\begin{aligned}
I_\mathrm{ext,cl} \approx  & I_e \int^{\inf}_{-\inf} \mathcal{N}(k|Np,Np(1-p))*\exp(-k\bar{\tau}_{\rm{cl}}){\rm{d}}k
\\
= & I_e \exp (-\frac{Np\bar{\tau}_{\rm{cl}}(2-\bar{\tau}_{\rm{cl}}+p\bar{\tau}_{\rm{cl}})}{2})\,.
\end{aligned}
\label{fulxgau}
\end{equation}
That is to say, the equivalent optical depth of foreground clumpy regions can be approximated by:

\begin{equation}
\hat{\tau}_{\rm{cl}}\approx\frac{Np\bar{\tau}_{\rm{cl}}(2-\bar{\tau}_{\rm{cl}}+p\bar{\tau}_{\rm{cl}})}{2} \,.
\label{equtau}
\end{equation}
Also, as $p <<1$ and if $\bar{\tau}_{\rm{cl}}<1$, the term $p\bar{\tau}_{\rm{cl}}$ in Equation \ref{equtau} can be neglected and $Np$ can be written as $n_{\rm{g}}{\pi}R^2_{\rm{cl}}$. Thus, we finally get
\begin{equation}
\hat{\tau}_{\rm{cl}}\approx \frac{n_{\rm{g}}{\pi}R^2_{\rm{cl}}\bar{\tau}_{\rm{cl}}(2-\bar{\tau}_{\rm{cl}})}{2} \,.
\label{equtau1}
\end{equation}
This final approximation shows that $\hat{\tau}_{\rm{cl}}$ is determined by the size $R_{\rm{cl}}$, the column number density $n_{\rm{g}}$ and the mean optical depth $\bar{\tau}_{\rm{cl}}$ of an individual foreground clump.

During the derivation of Equation \ref{fluxBio}, we have assumed $p<<1, N>>1$ and $\bar{\tau}_{\rm{cl}}<1$.  In Appendix \ref{continuum approximation}, we provide a detailed comparison of the approximation of Equation \ref{equtau}  to the numerical solutions of  Equation  \ref{fluxBio} for different parameter sets, where we show that Equation \ref{equtau1} only has a difference from the  numerical solutions at the level of $\sim 10\%$ for $\bar{\tau}_{\rm{cl}}<1$. As we will show in Section \ref{Modelling results}, our best model estimation indeed have $\bar{\tau}_{\rm{cl}}<1$.

We remind that the continuous approximation of Equation \ref{fluxBio} is only for solving Equation \ref{fulxgau}, not that the distribution of these star forming regions is itself continuous. Actually, the dust extinction effect of these clumpy regions is  different from that of the continuously distributed dust. We show such a comparison in Appendix \ref{sec:diff of cont and cl}. 

\subsection{Dust Extinction on Balmer Emission}
\label{Dust extinction on Balmer emission}

With the geometric models of the clumpy nebular disk (Equation \ref{rho_gas}  and \ref{equtau1}) and continuous ISM distribution (Equation \ref{rho_star}), we model the overall dust  extinction effect on a specific nebular region. For a given nebular cloud in a model galaxy along the line of sight to galaxy center, we parameterize its position with central distance $l$ and disk inclination $\theta$ and get  $r=l \sin\theta, h=l \cos\theta$, where $l$ takes positive and negative values at  the  proximal and distal ends respectively. 

For a given nebular region, its nebular emission is extincted by the dust of itself, foreground diffuse ISM dust, and other foreground clumpy regions. The total line of sight optical depth then is
\begin{equation}
\tau'_{\rm{g}}(l,\theta)=\tau_{\rm{cl}}+\tau_{\rm{s}}(l,\theta)+\hat{\tau}_{\rm{cl}}(l,\theta)\,
\label{tau'g}
\end{equation}
where $\tau_{\rm{cl}}$, $\tau_{\rm{s}}$  and $\hat{\tau}_{\rm{cl}}$  are the optical depths of the clumpy region itself, the foreground diffuse ISM dust, and other foreground clumpy regions, respectively. 

For the double exponential diffuse ISM dust disk parameterized by $h_{\rm{s}}$ and $R_{\rm{s}}$ (Equation \ref{rho_star}), the optical depth of the foreground ISM dust along the line of sight at $(l,\theta)$ is

\begin{equation}
\tau_{\rm{s}}(l,\theta)=
\begin{cases}
	0.5\tau_{\rm{s,A}}(\theta)(2-\exp(\frac{l}{0.5\tau_{\rm{s,A}}(\theta)}));  \,  for \,  l \leq 0\\
    0.5\tau_{\rm{s,A}}(\theta) \exp(-\frac{l}{0.5\tau_{\rm{s,A}}(\theta)});   \,   for\,  l > 0
    \end{cases},
    \label{taus}
\end{equation}
where $\tau_{\rm{s,A}}$ is the total optical depth along the central line of sight which follows Equation \ref{exp_tau}.

For $\hat{\tau}_{\rm{cl}}$, according to Equation \ref{equtau1}, $n_{\rm{g}}$ is the only parameter being function of $(l,\theta)$. With the global exponential disk distribution model of the clumpy regions (parameterized by $h_{\rm{g}},R_{\rm{g}}$), we have 
\begin{equation}
n_{\rm{g}}(l,\theta)=
\begin{cases}
	0.5n_{\rm{g,A}}(\theta)(2-\exp(\frac{l}{0.5n_{\rm{g,A}}(\theta)}));  \,  {\rm for} \,  l \leq 0\\
    0.5n_{\rm{g,A}}(\theta) \exp(-\frac{l}{0.5n_{\rm{g,A}}(\theta)});   \,   {\rm for}\,  l > 0
    \end{cases}  \,,
    \label{num_g}
\end{equation}
where $n_{\rm{g,A}}(\theta)$ is the total column number density of clumpy regions along the central line of sight:
\begin{equation}
n_{\rm{g,A}}(\theta)=2\rho_{\rm{g,0}}R_{\rm{g}}F_{\rm{g}}(\theta).
\label{n_gas}
\end{equation}

\subsection{Dust Extinction on Stellar Emission}
\label{Dust extinction  on  stellar emission}

Similar to Equation \ref{tau'g}, the line of sight optical depth to a stellar emission region at ($l,\theta$) is contributed by the foreground diffuse ISM dust and clumpy dust, 
\begin{equation}
\tau'_{\rm{s}}(l,\theta)=
\tau_{\rm{s}}(l,\theta)+\hat{\tau}_{\rm{cl}}(l,\theta)\,.
\label{tau's}
\end{equation}
Obviously, $\tau_{\rm{s}}(l,\theta)$ and $\hat{\tau}_{\rm{cl}}(l,\theta)$ follow Equations \ref{taus} and \ref{equtau1}, respectively.

    \subsection{Off-axis Effect }
    \label{Off-axis effect}
In Sections \ref{Dust extinction on Balmer emission} and \ref{Dust extinction on stellar emission}, we have parameterized  the foreground dust extinction on the stellar and nebular emissions at a specific region $(l,\theta)$ along the galaxy central line of sight. However, the SDSS fiber has an aperture of $1.5$ arcsec instead of being an area-free point toward the center of each galaxy. Therefore, there is a certain fraction of photons not along the line of sight to galaxy center, which also enter the fiber. Because the SDSS fiber aperture is smaller than the radius of a typical disk galaxy in our sample, the on-axis (line of sight to galaxy center) assumption is a good approximation when galaxies are viewed in face-on cases. However, when galaxies are viewed edge-on, the scale-height of our sample galaxies becomes comparable or even smaller than the SDSS fiber aperture. In this case, the off-axis effect (line of sight not along the galaxy center), as we show next, will cause significant biases.

We assume that the scale-height of the nebular emission regions $h_{\rm{g}}$ is smaller than that of the stellar disk $h_{\rm{s}}$ and both scale-heights are smaller than the SDSS aperture size. In this case, as shown by the schema figure (top panel of Figure \ref{CCC_model}), along the line of sight to the galaxy center, the stellar emissions suffer from the dust extinction effects of both of the clumpy nebular dust and diffuse ISM dust. While for the stellar emissions off the disk plane, the dust extinction from nebular regions becomes negligible.

To approximate and correct for this off-axis effect, we assume that there are a fraction of photons $(f_{\rm{off}})$ emitted from the stellar disk being off-axis, which is defined as:
\begin{equation}
 f_{\rm{off}}=\frac{h_{\rm{s}}-h_{\rm{g}}}{h_{\rm{s}}}.   
\end{equation}
These off-axis stellar emissions are not extincted by the nebular dust. On the other hand, the emissions from the nebular regions obviously do not have this off-axis effect.  

We consider a case where the disk inclination $\theta$ is relatively large but not completely edge-on. In this case, the effective disk height in the line of sight is expressed as 

\begin{equation}
    h'_{\rm{comp}}(\theta)=h_{\rm{comp}}\sin\theta +R_{\rm{comp}}\cos\theta \,.
    \label{heff}
\end{equation}
As we will show later, the nebular disk has larger scale-length  than that of the stellar disk. As a result, with the decreasing of disk inclination $\theta$, the effective height of nebular disk ($h'_{\rm{g}}$) starts to approach that of the stellar disk ($h'_{\rm{s}}$). We write the specific inclination when $h'_{\rm{g}}=h'_{\rm{s}}$ as $\theta_{\rm{crit}}$. We assume that there is no off-axis effect on the stellar  disk when $\theta<\theta_{\rm{crit}}$. Actually, when disk galaxy becomes face-on, both $h'_{\rm{s}}$ and $h'_{\rm{g}}$ will be larger than the SDSS fiber aperture, and thus there is no off-axis effect on both disks. 

In summary, because of the off-axis effect for the highly inclined disk galaxies, inside the SDSS fiber, a fraction of off-axis photons emitted from the thicker disk have not been extincted by the thinner component. In specific, this fraction is parameters by 

\begin{equation}
f_{\rm{off}}(\theta)=
\begin{cases}
	0;   \,  {\rm for} \,  \theta<\theta_{\rm{crit}}\\
    \frac{h'_{\rm{s}}-h'_{\rm{g}}}{h'_{\rm{s}}};    \,  {\rm for} \,  \theta \ge \theta_{\rm{crit}}
    \end{cases}  \,.
\end{equation}
with above modeling, the dust extincted stellar intensity at given position $(l,\theta)$ is written as
\begin{equation}
\begin{aligned}
&I'_{\rm{s}}(l,\theta)=I_{\rm{s}}\rho_{\rm{s}}(l\sin\theta,l\cos\theta)* \\
&[f_{\rm{off}}(\theta)e^{-\tau_{\rm{s}}(l,\theta)} +(\ 1-f_{\rm{off}}(\theta))e^{-\tau'_{\rm{s}}(l,\theta)}],
\label{ext_is}
\end{aligned}
\end{equation}
where $\tau_{\rm{s}}$ and $\tau'_{\rm{s}}$ are given by Equations \ref{taus} and \ref{tau's} respectively and $I_{\rm{s}}$ represents the stellar emissivity in arbitrary unit. For nebular emission, there is no off-axis effect. Therefore, the dust extincted intensity is
\begin{equation}
I'_{\rm{g}}(l,\theta)=I_{\rm{g}}\rho_{\rm{g}}(l\sin\theta,l\cos\theta)e^{-\tau'_{\rm{g}}(l,\theta)} \,,
\label{ext_ig}
\end{equation}
where $\tau'_{\rm{g}}$ is given by Equation \ref{tau'g}.

\subsection{Global Dust Attenuation}

In the above sections, we have modelled the dust extinction effect on both nebular emission and stellar emission for a given point $(l,\theta)$ in model galaxy along the line of sight. Next, we model the global dust attenuation effect by integrating the dust extincted radiation along the line of sight while neglecting the dust scattering effect,
\begin{equation}
I_{\rm{comp,A}}(\theta)=\int_{-inf}^{inf}I'_{\rm{comp}}(l,\theta){\rm{d}}l .
\label{flux_att}
\end{equation}

For a given inclination $\theta$ of a disk galaxy, the unreddened intensity along the line of sight to galaxy center is

\begin{equation}
\begin{aligned}
I_{\rm{comp,*}}(\theta)=&\int_{-inf}^{inf}I_{\rm{comp}}\rho_{\rm{comp}}(l\sin\theta,l\cos\theta) {\rm{d}}l 
\\
=& 2I_{\rm{comp}}\rho_{comp,0}R_{\rm{comp}}F_{\rm{comp}}(\theta) \,.
\label{flux_un}
 \end{aligned}
\end{equation}
In Equation \ref{flux_att} and Equation  \ref{flux_un}, the subscript $`\rm{comp}'$ refers to $`\rm{s}'$ for stellar disk and $`\rm{g}'$ for nebular disk. With the unattenuated and attenuated nebular emission line intensity, the dust attenuation then is easily defined as 
\begin{equation}
A_{\rm{comp}}=-2.5 \log (\frac{I_{\rm{comp,A}}}{I_{\rm{comp,*}}})
\label{Alam}
\end{equation} 
and so that the reddening is $E_{\rm{comp}}(\lambda_1-\lambda_2)=A_{\lambda_1,\rm{comp}}-A_{\lambda_2,\rm{comp}}$. In  above equations, we have assumed the dust extinction curves of both clumpy nebular and diffuse ISM dust follow the same simple power-law (Equation \ref{power curve}).
    \label{Global dust attenuation}

    \subsection{MCMC Fitting}
    \label{MCMC fitting}
        \subsubsection{Reducing free parameters}
        \label{Reducing free parameters}
        
In the CCC modeling, we have used 11 parameters in above equations, which are listed in Table \ref{2com}. For modelling the stellar disk, we have used $h_{\rm{s}}$, $R_{\rm{s}}$, $\alpha_{\rm{s,0}}$ , $I_{\rm{s}}$ (Equation \ref{exp_tau}, \ref{taus}, \ref{tau's}, and \ref{ext_is}). For the nebular disk, we have used $h_{\rm{g}}$, $R_{\rm{g}}$,  $\rho_{\rm{g,0}}$, $I_{\rm{g}}$, $\tau_{\rm{cl}}$, $R_{\rm{cl}}$, $R_{\rm{a}}$ (Equation \ref{equtau1}, \ref{num_g}, and \ref{ext_ig}), where we remind that $\tau_{\rm{cl}}$ is  at the wavelength 5,500 $\text{\AA}$ and is directly related to the effective optical depth $\bar{\tau}_{\rm{cl}}$ through Equation \ref{eff_tau}. Since the dust attenuation describes the ratio of the emission with and without dust, the emissivity $I_{\rm{s}}$ and $I_{\rm{g}}$  are therefore unnecessary parameters. For simplicity, we set $I_{\rm{s}}$ and $I_{\rm{g}}$ to be $1$.  

Then, the dust attenuation value at a given inclination is determined by two sets of parameters: the geometric  parameters:$h_{\rm{g}}$, $R_{\rm{g}}$, $h_{\rm{s}}$, $R_{\rm{s}}$, $R_{\rm{a}}$, $R_{\rm{cl}}$ and dust-related parameters: $\alpha_{\rm{s,0}}$, $\rho_{\rm{g,0}}$, $\tau_{\rm{cl}}$. For these geometric parameters, their absolute values are actually degenerated with the central density parameter (e.g. Equation \ref{exp_tau}, \ref{taus}, and \ref{n_gas}). That is to say, the absolute sizes of our model galaxies can be varied arbitrarily, provided that their corresponding density parameters are adjusted accordingly. Therefore, following Section \ref{e_s mix}, we normalize these geometric parameter sets by setting the stellar  scale-length $R_{\rm{s}}=1$. Then, the rest scale parameters $h_{\rm{g}}$, $R_{\rm{g}}$, $h_{\rm{s}}$ becomes equivalent scale-length and scale-height in unit of $R_{\rm{s}}$. The corresponding density parameter $\alpha_{\rm{s,0}}$ and $\rho_{\rm{g,0}}$ become equivalent densities in unit of $R_{\rm{s}}^{-1}$. 

In addition, during the modeling of the dust extinction effect of the clumpy regions, the size $R_{\rm{cl}}$ of the clumps is coupled with their column number density (Equation \ref{equtau1}).  Therefore, in our final modeling of global dust attenuation effect, $R_{\rm{cl}}$ is coupled with the central number density $\rho_{\rm{g,0}}$. Considering this coupling effect, we define a new parameter $\sigma
_{\rm{g,0}}\equiv\rho_{\rm{g,0}}{\pi}R^2_{\rm{cl}}$, which represents the cross-section of the clumpy regions in galaxy center.

In summary, our model has 6 independent free parameters to be constrained. Among them, there are 3 equivalent scale-height and scale-length parameters: $h_{\rm{g}}$, $R_{\rm{g}}$, and $h_{\rm{s}}$; central absorption density of diffuse ISM dust $\alpha_{\rm{s,0}}$; central cross-section $\sigma_{\rm{g,0}}$ of the clumpy regions and the  optical depth of an individual clumpy region $\tau_{\rm{cl}}$. 

\subsubsection{Parameter constraints }
\label{Parameter constraints}

We use CCC model to fit the observed inclination dependence of $E_{\rm{g}}(\theta)$ and $E_{\rm{s}}(\theta)$ simultaneously, where a MCMC algorithm is used to make the best estimates of six free parameters. The probe of the model parameters is set in logarithmic space. Also, in order to obtain physically meaningful parameter values, we set the following constraints on the model parameters.

$\bullet$ Both nebular and stellar disk shall have a larger scale-length than scale-height: $R_{\rm{s}}> h_{\rm{s}}, R_{\rm{g}}>h_{\rm{g}}$;

$\bullet$ The scale-length of the two disks shall be comparable: $0.5 < R_{\rm{s}}/R_{\rm{g}} < 2 $;

$\bullet$ The scale-height of nebular disk shall  be smaller than that of stellar disk: $h_{\rm{g}} < h_{\rm{s}}$;

$\bullet$ The optical depth of SF regions shall follow the optically thin approximation: $\tau_{\rm{cl}}<1$;

\subsubsection{Modelling results}
\label{Modelling results}

After excluding invalid MCMC chains, we get excellent fittings to the observed $E_{\rm{g}}-\theta$ and $E_{\rm{s}}-\theta$ relations and strong constraints on all six model parameters. We show the corner plot of the MCMC fitting in  the bottom panel of Figure \ref{CCC_model} and list the fitting results in Table \ref{2com}. As can be seen, the CCC model fits the $E_{\rm{g}}-\theta$ relation of these highly inclined disks ($\theta > 70^\circ$) very well, where the simple screen model  has failed, as described in Section \ref{e_g screen}.
  
 For the diffuse ISM component, we get the best fits that the height-to-length ratio $h_{\rm{s}}/R_{\rm{s}}= 0.19$ and the central absorption  density of ISM dust $\alpha_{\rm{s,0}} = 2.58$.  For CCC model, using Equation \ref{exp_tau}, we can easily estimate that the full optical depth of the diffuse ISM component along the galaxy center increases from $\tau_{\rm{s}}(0^\circ)=0.95$ with face-on inclination to $\tau_{\rm{s}}(90^\circ)=5.16$ with edge-on inclination. We note that the height-to-length ratio  in our model results is slightly larger than that of  reported in literature, which is around 0.1-0.15 \citep{Shao2007,Guthrie1992}.  The specific reasons for this discrepancy will be discussed in sections \ref{Dust geometry} and \ref{caveats}. For the global distribution of nebular emission line regions, the CCC model presents a very thin clumpy nebular disk, which has a height-to-radius ratio,$h_{\rm{g}}/R_{\rm{g}}= 0.066$, and is much thinner than that of diffuse ISM disk.  More specifically, in  our modelling, the scale-length of the nebular disk is about 1.6 times larger than the diffuse ISM disk, while the scale-height is only about half of the diffuse ISM dust disk ($h_{\rm{g}}/h_{\rm{s}}= 0.55$). The decrease of the Balmer decrement at high inclination requires that the nebular clumpy disk is more extended than the diffuse ISM dust component,  such that the outer disk clumps are less attenuated by ISM dust .

For individual clumpy regions, our model shows that the number of clump regions at galaxy center per stellar scale-length $R_{\rm{s}}$  ($\sigma_{\rm{g,0}}\equiv \rho_{\rm{g,0}}\pi R_{\rm{cl}}^2$) is about 1.8.  Therefore, we expect that there are only about 0.37 clumps along the central line of sight when our model galaxy is face-on ( Equation \ref{n_gas}). This number increases to 5.61 when the model galaxy is edge-on, which means  an overlapping effect of HII regions. That is right the reason of the failure of the simple screen model at high inclinations presented in Section \ref{e_g screen}, where no overlapping effect has been assumed. 

For the optical depth of individual emission line region in the CCC model, we get the best estimate of  $\tau_{\rm{cl}}= 0.5$, which is very close to the result we have obtained in the simple screen model (Section \ref{e_g screen}). This result is not surprising. When the model galaxy is face-on, as we have  discussed, there is essentially few overlapping effect of the HII regions along the line of sight, and the only extra dust extinction in addition to the ISM dust that affects an emission line region is only from the local dusty shell of this region itself.  In this case, the CCC model degenerates to the dust screen model in Section \ref{e_g screen}. 

Besides, in our modelling, the critical viewing angle where the effective scale-height of nebular disk equals to that of diffuse ISM disk happens at $\theta_{\rm{crit}}\approx 83^{\circ} $. That is to say, only for these very highly inclined disk, we have considered the off-axis effect for the nebular dust extinction observed by the SDSS fibers (Section \ref{Off-axis effect}).

\section{APERTURE EFFECT AND DUST ATTENUATION}
\label{APERTURE EFFECT AND DUST ATTENUATION}

As we have shown in Figure \ref{egest} and discussed in Section \ref{e_g screen}, when a galaxy disk is highly inclined, there is a saturation effect on the observed dust reddening of the nebular emission lines, which is caused by the fact that the inner emissions would be nearly completely extincted by the outer dust layers along the highly inclined line of sight. That is to say, in observation, we can not distinguish whether such a saturation effect has occurred or not from color excess values alone. On the other hand, if we further consider the extincted flux, we would easily find how much of the emission flux has been extincted by dust and then distinguish the saturation effect of color excess.

\begin{figure}[htbp]
\centering
\gridline{\fig{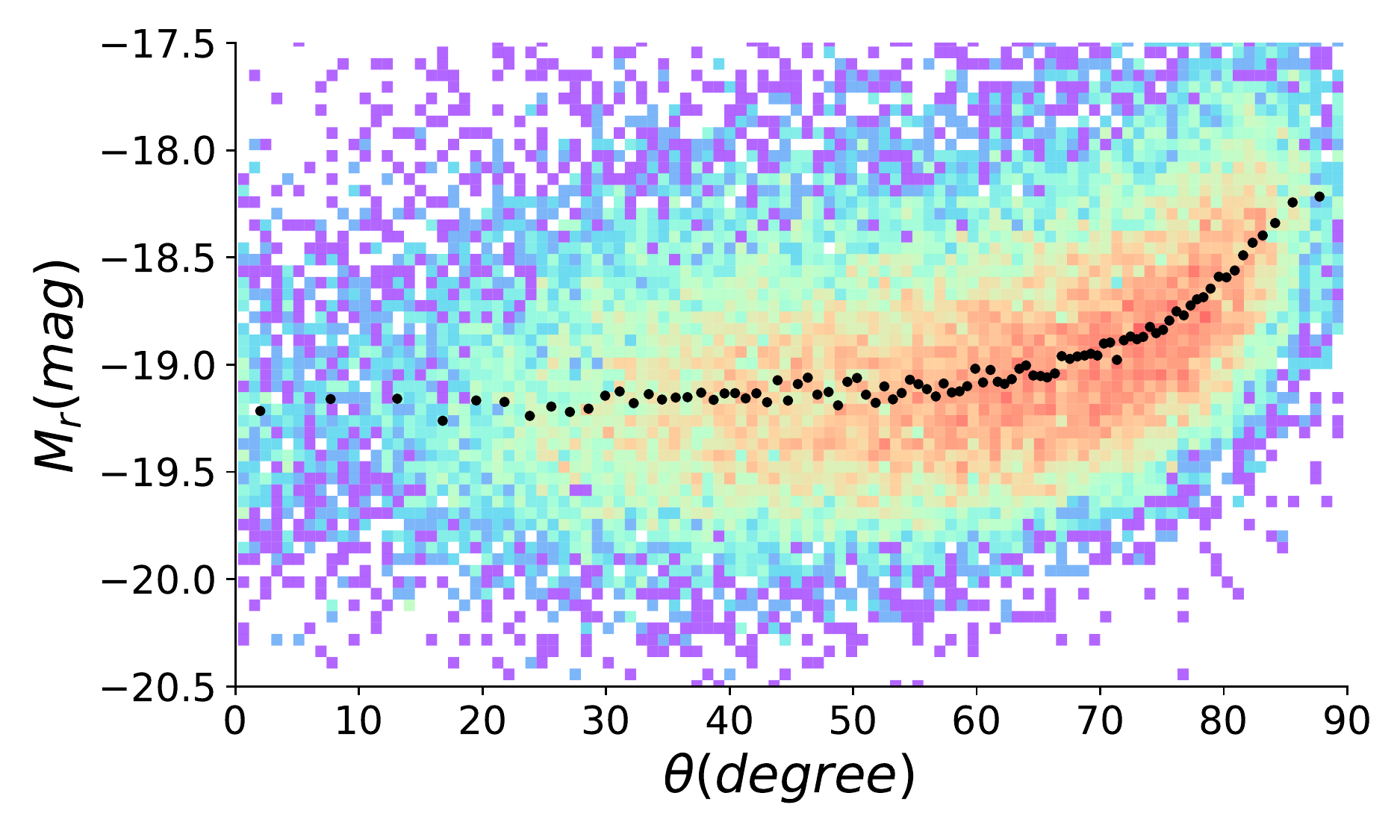}{3in}{(a) $M_r-\theta$ distribution}
          }
\gridline{\fig{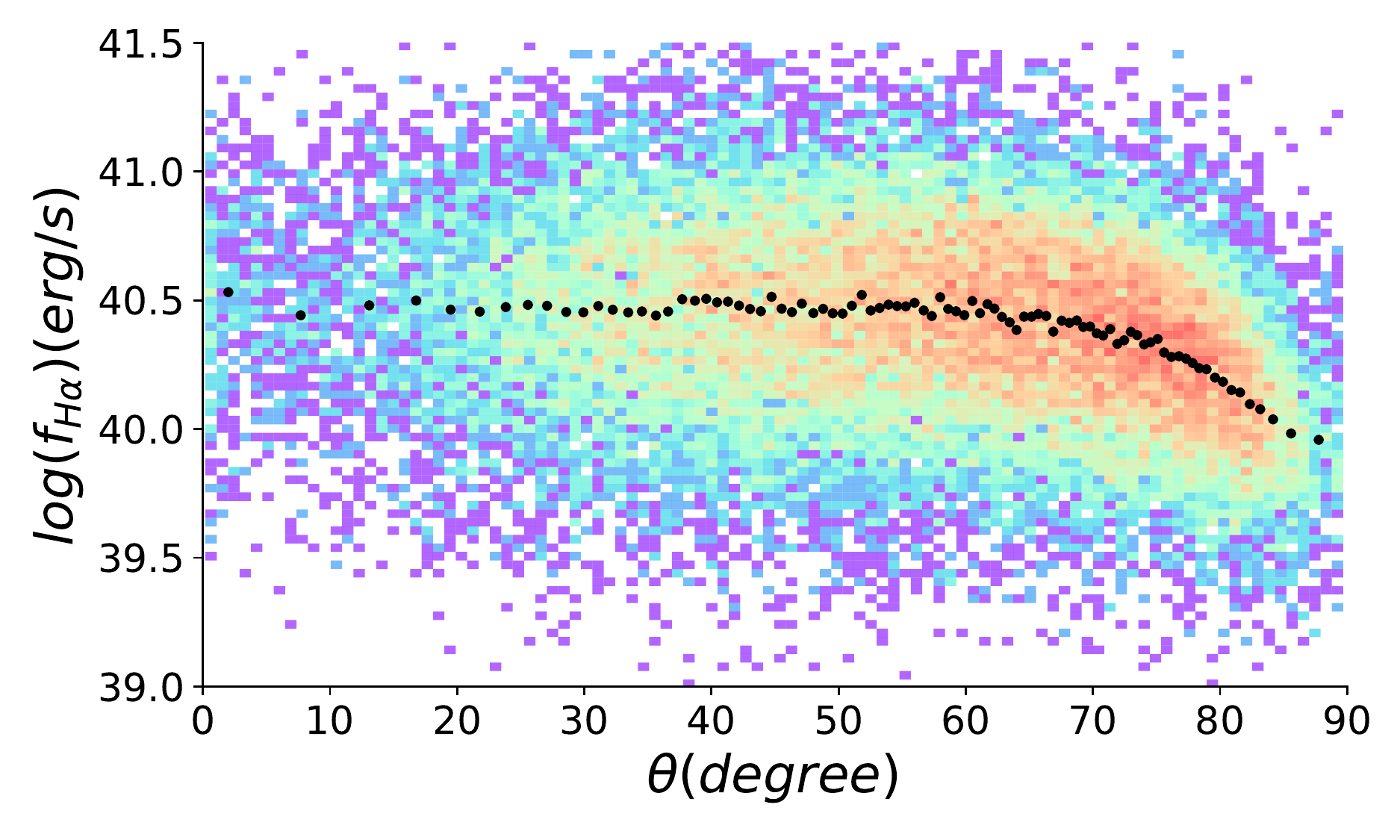}{3in}{(b) $F_{H\alpha}-\theta$ distribution}
          }

    \caption{ The absolute $r$ band fiber magnitude(top panel)  and $\rm{H}_\alpha$flux(bottom panel) as function of disk inclination $\theta$ for our MW-like SFGs, where the number density of the sample galaxies in the parameter space are color coded. The median values at different inclinations are shown as solid dots in each panel.}
    \label{flux_theta}
 \end{figure}

We show the inclination dependence of the $r$ band absolute magnitude and  $\rm{H}\alpha$  flux inside fiber aperture of our sample galaxies in Figure \ref{flux_theta}.  At given inclination, the median absolute fiber magnitude and  $\rm{H}\alpha$  flux are shown by the solid dots, which  can be viewed as the inclination dependence of the observed fiber magnitude and  $\rm{H}\alpha$  flux for a typical MW-like SFG at typical redshift $z\sim0.07$.  As can be seen, both of the fiber magnitude and   $\rm{H}\alpha$ flux  do not change when the disk inclination is low ($\theta < \sim 50^\circ$) and then increases/decreases  monotonically with increasing disk inclination for the fiber magnitude and   $\rm{H}\alpha$ flux respectively. Overall, the fiber magnitude dims about 1 mag and the  $\rm{H}\alpha$ flux drops about 0.5 dex from the face-on view to edge-on view.

\begin{figure}[htbp]
   \centering   
    \includegraphics[width=3in]{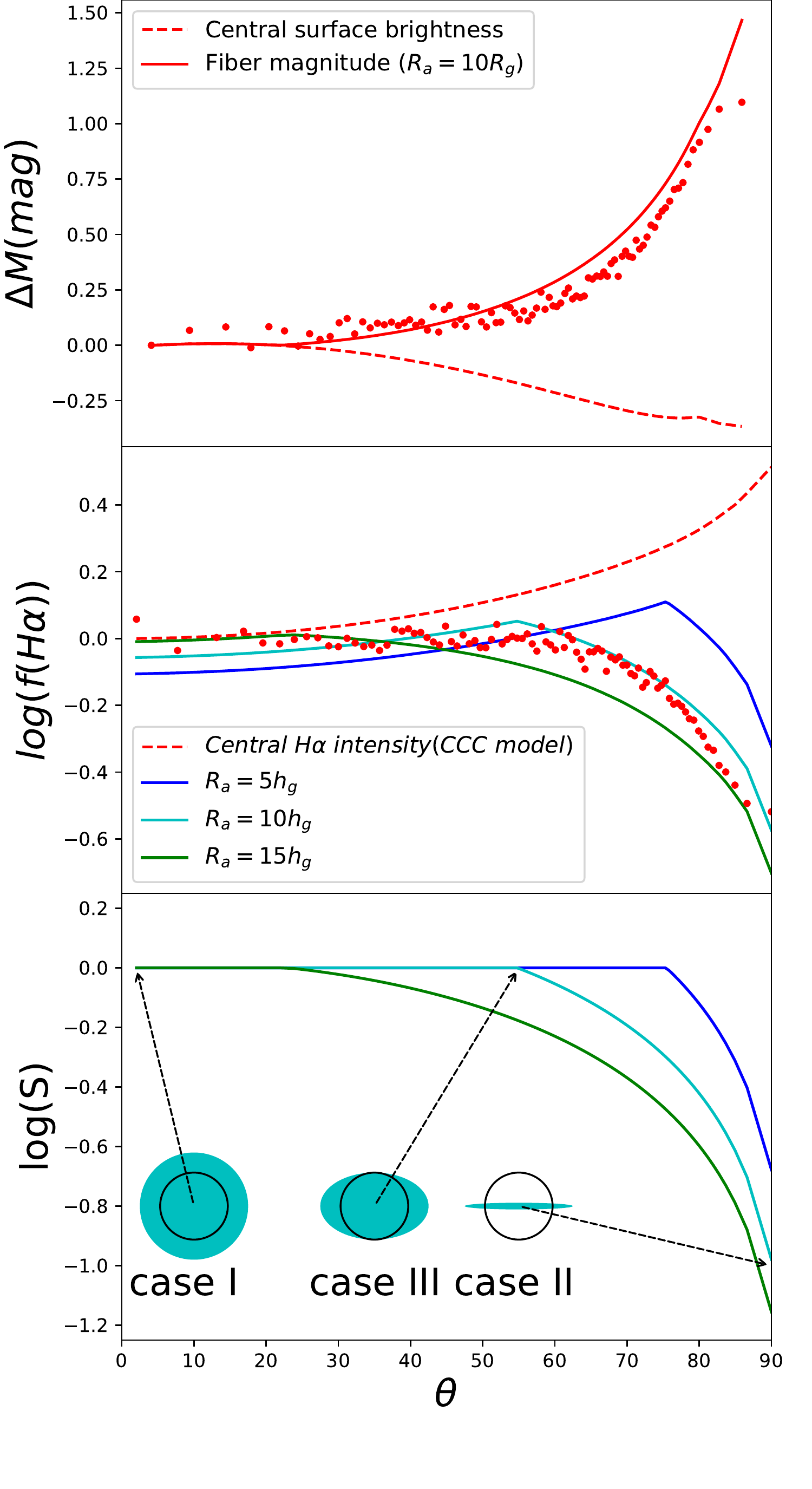}
    \caption{ The inclination dependent $\rm{H}\alpha$ flux and $r$ band fiber magnitude explained by the CCC model and aperture effect. Top panel: the observed median fiber magnitude in $\theta$ bins (dots), the attenuated central surface brightness predicted by the CCC model (dashed line), the aperture corrected fiber magnitude  after applying aperture corrections for the model galaxy with $R_{\rm a}/h_{\rm g}=10$(solid line).   Middle panel: the observed median $\rm{H}\alpha$  flux in $\theta$ bins (dots), the attenuated central nebular  line fluxes predicted by the CCC model (dashed line), the aperture corrected nebular  line fluxes after applying aperture corrections for three model galaxies with different physical sizes (solid lines, different colors represent different $h_{\rm{g}}$ values). Bottom panel: the projection of a double-exponential disk into a fiber with radius $R_{\rm{a}}$ (2.2 kpc) at different inclinations. The three solid lines show the projected area $S$ as function of disk inclinations for three model galaxies, which are all set to unit value at $\theta=0^\circ$. The three schematic icons show the projection of a model galaxy with scale-length 3.2 kpc and scale-height 0.2 kpc into the fiber aperture (2.2 kpc) for face-on (case I), edge-on (case II) and critical (case III) inclinations respectively.}
    \label{apcorr}
 \end{figure}

We  use the best CCC model to calculate the attenuated nebular emission line flux densities  and $r$ band stellar  surface brightness along the central line of sight  for different $\theta$ values. The output (attenuated)  $r$ band stellar surface brightness and $\rm{H}\alpha$  flux density are plotted as the dashed curves in the top and middle panels of Figure \ref{apcorr}. The emission line flux and surface brightness have been both normalized to zero at $\theta= 0^\circ$. When dust attenuation effect is considered,  the attenuated central flux density from the CCC model prediction generally shows a plateau at low inclination  ($\theta < \sim 50^\circ$), which is consistent with observations. While for high inclinations  ($\theta > \sim 50^\circ$), the CCC model predicts a increasing central flux density with disk inclination, which apparently is contrary to the decreased trend of the aperture flux seen by the observational data.

The inconsistency between the increasing trend of the central  flux densities and the decreasing trend of the aperture fluxes can be explained by the aperture effect. In other words, what we observe is the total flux inside the fiber aperture, while our model prediction is the flux density along the central line of sight. To estimate the observed  flux correctly, we need to take the aperture effect into account.

For our sample galaxies, the physical aperture size of SDSS fibers ($R_{\rm{a}}\sim 2.2$ kpc)  brings important differences between the observed flux inside a fiber and the central line of sight flux density in our model. On one hand, the projected nebular emission line (or stellar continuum) flux inside a fiber is not a constant, but has a profile depending on both the disk inclination and dust geometry in a complicated way. On the other hand, when the disk is highly inclined, the projected disk height (especially that of clumpy nebular disk) will not fill all of the fiber aperture.  We show the fiber aperture effect for galaxies with different inclinations (three cases) with schematic icons in the bottom panel of Figure \ref{apcorr}. 

Detailed analytical modeling of the projected emission line and stellar continuum fluxes inside fiber aperture as a function of disk inclination is highly complex and beyond the scope of this study. Here, we make approximations based on simplified assumptions. For brevity,  we illustrate below the process of estimating the emission line flux as function of different inclinations. For the stellar emission ($r$ band fiber magnitude), the estimation process can be referred exactly to that of the emission lines.

We assume that the observed nebular emission line flux is proportional to the product of the central surface brightness $I_{\rm{cen}}$ and disk projected area in aperture $S$,
\begin{equation}
    L_{comp}(\theta,R_{\rm{a}})= I_{\rm{cen},comp}*S,
    \label{apcorrL}
   \end{equation} With CCC model, we have derived the central flux density $I_{\rm{cen},comp}$ as a function of disk inclination (dashed line in middle panel of Figure \ref{apcorr}). Next, we discuss the values of $S$ at different inclinations. 

For a MW-like disk galaxy with stellar mass $10^{10.4}M_{\odot}$, the scale-length of stellar disk is about 2 kpc \citep{Shen2003}. Adopting this value as the scale-length of stellar disk $R_{\rm{s}}$ in CCC model, we obtain that $h_{\rm{s}},R_{\rm{g}},h_{\rm{g}}$ of the  disk of our model galaxy are 0.37 kpc, 3.17 kpc and 0.21 kpc respectively. When the model galaxy is face-on, the emission line disk scale-length is larger than the fiber aperture,  the projected disk will fill all of the fiber aperture and thus $S=\pi R_{\rm{a}}^2$ (case I in Figure \ref{apcorr}). When the model galaxy is  edge-on, the projected area is determined by the scale-height of the disk.  In this case, we approximate the projection area as $S\approx\pi R_{\rm{a}}h_{\rm{g}}$ (case II  in Figure \ref{apcorr}). For a disk with an arbitrary inclination $\theta$, following Equation \ref{heff} of Section \ref{Off-axis effect}, its equivalent projection scale-height can be written as  $h_{\rm{comp}}' (\theta)=h_{\rm{comp}}\sin\theta +R_{\rm{comp}}\cos\theta $ as the approximation of the projected height of disk,  we can consistently write $S$ as: 

\begin{equation}
S(\theta) \approx\pi R_{\rm{a}} {\rm Min} (h_{\rm{comp}}'(\theta),R_{\rm{a}}),
\label{sarea}
\end{equation}
  where  '$\rm Min$' takes the minimum value between $R_{\rm{a}}$ and $h'_{\rm{comp}}(\theta)$. With this approximation, there is a critical inclination angle $\theta'_{\rm{crit}}$ where $h_{\rm{comp}}'(\theta'_{\rm{crit}})=R_{\rm{a}}$ (case III in Figure \ref{apcorr}).  When the model galaxy is observed from face-on to critical inclination $\theta'_{\rm{crit}}$ ($\theta < \theta'_{\rm{crit}}$), we have $h_{\rm{comp}}' \geq R_{\rm{a}}$, and $S$ equals to the covering area of fiber, $\pi R^2_{\rm{a}}$. When $\theta > \theta'_{\rm{crit}}$,  $S = \pi R_{\rm{a}}h_{\rm{comp}}'$.

In the middle panel of  Figure \ref{apcorr}, we plot the model predicted $\rm{H}\alpha$ fluxes inside fibers as functions of disk inclinations for three different chosen $R_{\rm{a}}/h_{\rm{g}}$ values (5,10,15)  using solid lines of different colors. Since $R_{\rm{a}}$ has a physical size $\sim 2.2$ kpc,  these three different  $R_{\rm{a}}/h_{\rm{g}}$ ratios imply three different physical sizes of the model galaxies. Because the best estimate of the CCC model have $R_{\rm{g}}/h_{\rm{g}}=1.6$, for these three $R_{\rm{a}}/h_{\rm{g}}$ values (5,10,15), we have  $\theta'_{\rm{crit}} \sim 75 ^\circ, 55 ^\circ,25 ^\circ$ respectively. For each $R_{\rm{a}}/h_{\rm{g}}$ value, we then normalize the model-predicted mean $\rm{H}\alpha$ fluxes for inclinations between 0 and $\theta'_{\rm{crit}}$  to unit value.  When $R_{\rm{a}}/h_{\rm{g}} = 5$,   $\theta'_{\rm{crit}}\sim 75^\circ$, because the central flux density $I_{\rm{cen}}$ still increases in this inclination range (larger than 0.1 dex), our model does not predict a very flat plateau in this inclination range. Moreover, from  $\theta'_{\rm{crit}}$ to edge-on, the drop of the  model predicted $\rm{H}\alpha$ flux is also not as large as that being observed. On the other hand, when $R_{\rm{a}}/h_{\rm{g}} = 15$, because  then $R_{\rm{g}} < R_{\rm{a}}$, our model predicts a continuous decrease of the effective projection area (Equation \ref{sarea}) and resulted $\rm{H}\alpha$ flux, which looks also not be in good consistence with observations. Finally, we see that $R_{\rm{a}}/h_{\rm{g}} = 10$ provides a fairly good prediction on the global behavior of the observed nebular emission line flux as a function of disk inclination. In fact, we also have calculated the sum squared residual values of these three different model lines to the observational values in each $\theta$ bin, and find that  the line of $R_{\rm{a}}/h_{\rm{g}}= 10$ is indeed the best.  Moreover, adopting  $R_{\rm{a}}/h_{\rm{g}} = 10$  and replacing the parameters with those of the stellar disk,  we can make a similar aperture effect correction for the observed  $r$ band fiber magnitude using Equation \ref{apcorrL},\ref{sarea}. The result is shown as the solid line in the top panel of Figure \ref{apcorr}. Again, we see that, after correction of aperture effect,  our CCC model makes an excellent prediction on the inclination dependence of dust attenuation effect for $r$ band stellar continuum. It is worth emphasizing that the constraints on the parameters of our CCC model are from the  the reddening features of the emission lines and the stellar continuum,  without using their attenuation features. In the  above, we show that the CCC model predicted relations between the attenuation features and disk inclination are also very consistent with the observations once the fiber aperture effect is probably accounted. This result further illustrates the internal self-consistency of our CCC model in predicting the dust attenuation and reddening features of  local disk galaxies.

 The excellent consistence of $R_{\rm{a}}/h_{\rm{g}}=$ 10 model with observations provides an interesting constraint on the physical size of our model galaxy. In Section \ref{CCC model}, because of the degeneration of model parameters, we can only obtain the relative geometry parameters (all in unit of $R_{\rm{s}}$ ) for our model galaxy. Here, because of the average aperture size  $R_{\rm{a}}\sim 2.2$ kpc as we have discussed, we naturally obtain  $h_{\rm{g}}= 0.22$ kpc and then get estimates of all the  model parameters in physical units. We list them in the last column of Table \ref{2com}. With the physical sizes of all these geometric parameters, we can explore further the physical implications of the CCC model.

\section{DISCUSSION}
\label{DISCUSSION}
In Section \ref{CCC model}, we have obtained the best estimates of the geometric parameters of the CCC model in units of $R_{\rm{s}}$. Besides, we also get a constraint that each clumpy region has optical depth of $\sim 0.50$ in $V$ band. In Section \ref{APERTURE EFFECT AND DUST ATTENUATION}, we find that, by  using the aperture effect, we get estimates of the model parameters in physical units: $R_{\rm{s}}\sim 2.1$ kpc, $R_{\rm{g}}\sim 3.33$ kpc, $h_{\rm{s}}\sim0.41$ kpc, $h_{\rm{g}}\sim 0.22$ kpc, $\alpha_{\rm{s,0}}\sim 1.22$ kpc$^{-1}$, $\sigma_{\rm{g,0}}\sim 0.84$ kpc$^{-1}$ for modeled MW-like disk galaxies. In this section, we compare our model estimates with  other observational results of local massive disk galaxies and make further discussions.

\subsection{Dust Geometry}
    \label{Dust geometry}
    
We first compare the  geometry  parameter of our model galaxy with observational or modelling results of the Milky-Way and  other nearby disk galaxies.  

For the overall structural parameters of the dust component of the Milky-Way, many studies have reached consistent conclusions that dust is thinner and more extend than stars \citep[e.g.][]{Drimmel2001,Misiriotis2006,Li2018}.  For the extra-galactic galaxies, most of the studies on the dust geometry in optical wavelengths also have assumed only one global dust component and obtained similar conclusions. For example, \citet{Xilouris1999}(hereafter X99) investigated the surface brightness profiles of five nearby edge-on galaxies in  B, V, and I bands,  \citet{Bianchi2007} analyzed another seven nearby edge-on galaxies in  V and K bands. By applying radiative transfer analysis on these galaxies, they find that the radial scale-length of dust is about 1.4 times larger than that of  stars, while its vertical scale-height is about half of the stellar disk. More recently, \citet{DeGeyter2014} studied the SDSS $g, r, i,$ and $z$ band images of 12 edge-on spiral galaxies selected from the CALIFA survey and found similar conclusions. Besides  edge-on galaxies, \citet{Casasola2017} investigated 18 face-on spiral galaxies in DustPedia from UV to sub-millimeter bands and found that the dust scale-length is about 1.6 times of the stellar one. For our model galaxy, the clumpy nebular disk has a larger scale-length and smaller scale-height than the stellar disk, while the diffuse ISM dust component has been assumed to follow the same geometry as  the stellar disk. Qualitatively, merging the dust in the clumpy regions with that in the diffuse ISM  will give a global dust component that is larger in scale-length and smaller in scale-height than the stellar component, which is consistent with other studies. However, considering the non-linearity of the dust attenuation effect, the geometric parameters of different dust models are not quantitatively  comparable.

\begin{figure}[htbp]
    \centering
    \includegraphics[width=3in]{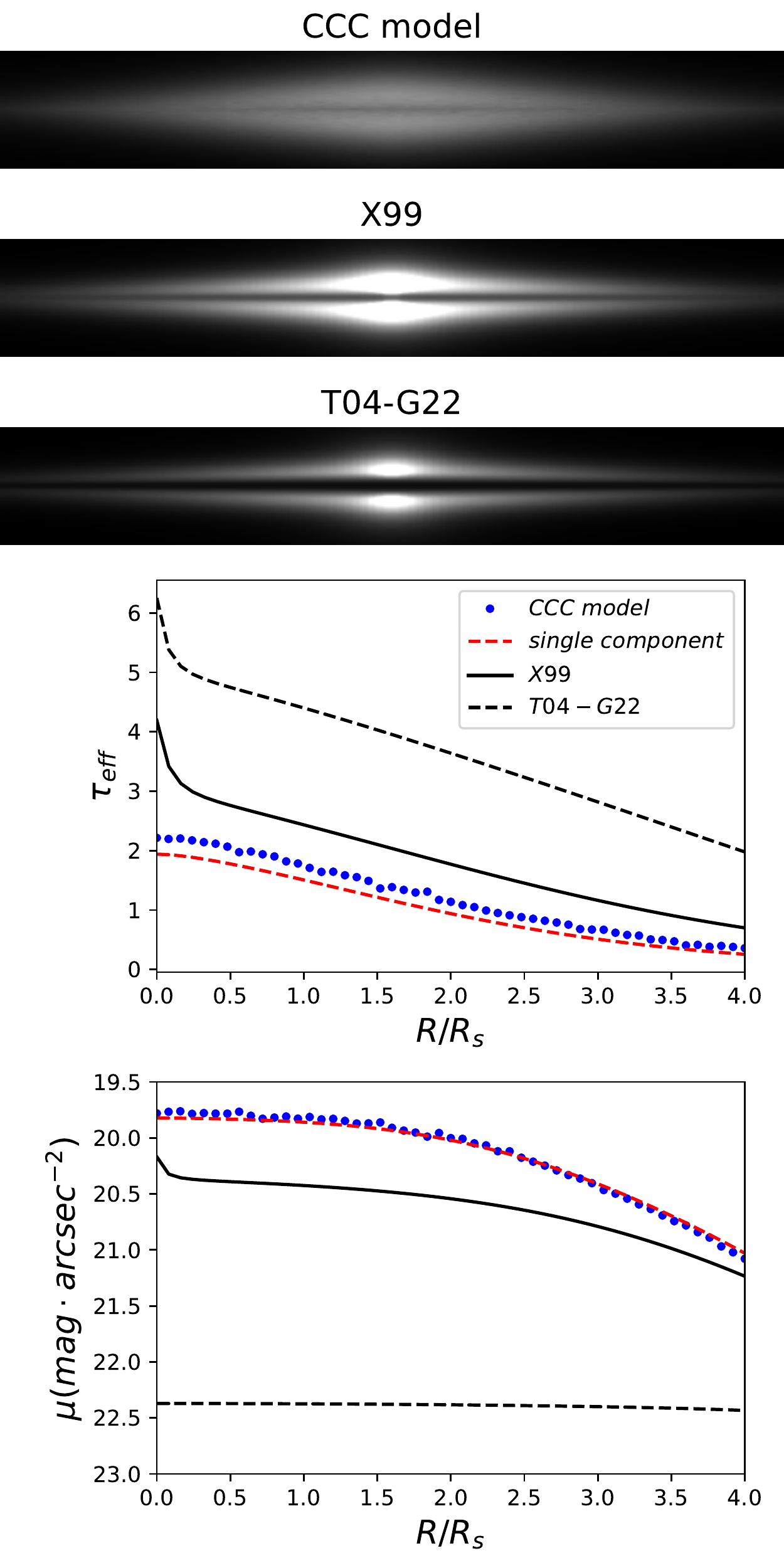}
    
    \caption{The edge-on projection images, effective optical depth and surface brightness profiles of the CCC, X99 and T04 model galaxies. The top three image panels: the  edge-on projection images of the CCC,X99 and T04 models. The middle plot panel: the effective optical depth profiles along the galactic plane of the three models. The bottom panel: the surface brightness profiles along the galactic plane of the three models.  }
\label{img}
 \end{figure}
 
To quantify the global dust attenuation effect of the two dust components in our CCC model, we reconstruct the projected image of our model galaxy from the edge-on view  using the best model parameters.  More specifically, the image projection process is an integral of the dust attenuated stellar emissions along the line of sight, which is given by
\begin{equation}
\begin{aligned}
    I_A(X,Y)=\int^{inf}_{-inf}I_0\rho_s(X,Y,l_{\rm{z}})  e^{-\tau(X,Y,l_z)}{\rm{d}}l_{\rm{z}}, \\
\end{aligned}
\label{projection}
\end{equation}
where $\rho_s$ and $\tau$ denote the density of stellar emission and dust optical depth at given position $(X,Y,l_z)$  respectively,  and $(X,Y,l_z)$ is a Cartesian coordinate system  with the centre of the model galaxy as the origin. $(X, Y)$ forms the projection plane and $X,Y$ is the major and minor axis along the disk plane,separately, while $l_z$ is the axis perpendicular  to the projection plane. This Cartesian coordinate system can be easily mapped to the cylindrical coordinate system $(r,h)$ that has been defined in Equation \ref{Dexp} through

\begin{equation}
\begin{aligned}
r&=\sqrt{X^2+l_z^2},
\\
h&=Y.
\end{aligned}
 \label{mapping}
\end{equation} 

As can be seen from the Equation \ref{projection}, the resulted image is not only a function of geometric parameters  (e.g. scale-length and scale-height, discussed in detail below), but also is a function of the normalization parameter  of dust optical depth (or density). In addition, depending on the model assumptions, $\rho_s(X,Y,l_z)$ and $\tau(X,Y,l_z)$ can both be combined by multiple components. For  example, $\rho_s$ of our CCC model is only  an exponential disk characterized by $R_s,h_s$, while $\tau$ is combined of two disks, the diffuse dust  (c.f. Equation \ref{exp_tau0}) and the clumpy dust (c.f. Equation \ref{equtau1}).

With above equations, the resulted  $V$-band edge-on model galaxy image of our best fit CCC modelling is shown in the top panel of  Figure \ref{img}. To do the projection, we have set the apparent magnitude of our model galaxy without any dust attenuation to be 17.5 mag and its stellar disk scale-length $R_s=1.5$ arcsec (typical values of our sample galaxy). In this reconstructed image, a dust lane structure along the middle plane of the model galaxy is clearly seen, which is originated from the extra obscuring effects of the clumpy dust component. To further quantify the global properties of the projected image, we plot the effective optical depth $\tau_{\rm eff}$\footnote{The effective dust optical depth is defined through  $I=I_0*{\rm exp}(-\tau_{\rm eff})$, where $I$ and $I_0$ are the observed surface brightness along the line of sight with and without dust attenuation respectively.} and the surface brightness profiles along the galactic mid-plane in the bottom two panels of Figure \ref{img} as dotted lines. With these two quantitative profiles, we make more detailed discussions on the dust geometry of our best fit CCC model and  further compare it with other dust attenuation models of disk galaxies.

\subsubsection{Equivalent single dust component model}
\label{equivalent single dust component}

Since many early studies of the dust geometry of disk galaxies considered only one dust component, we are interested in testing whether the two dust components in our CCC model can be equated with one dust component. To test this idea, we consider a simple model of a continuously distributed double exponential dust disk mixed with a stellar emission disk which is assumed to be the same as that of the CCC model. To predict the observed surface brightness profile for this single dust component model,  as that shown in the bottom  panel of Figure \ref{img},  three model parameters are needed: the dust-to-stellar scale-length ratio $R_d/R_s$, the  dust-to-stellar scale-height ratio $h_d/h_s$ and a normalization parameter representing the  dust density. For this dust density parameter, follow convention, we take the V-band the central optical depth of the model galaxy in face-on view $\tau_f$. By adjusting these three model parameters, we find that  $R_d/R_s \sim 1.1$, $h_d/h_s\sim 0.7$ and $\tau_{f}\sim  0.4$  provide almost the same surface brightness profile as that of the CCC model prediction, which is shown as the red dashed lines in the bottom two panels of Figure \ref{img}.  For this equivalent single dust component, the dust scale-length is  about $10\%$ larger than the stellar scale-length. This result is  in good agreement with that of \citet{Munoz-Mateos2009} for nearby disk galaxies, where the global dust scale-length is obtained from the modelling of the infrared emissions.

From Figure \ref{img}, we conclude that our two-component CCC model can indeed be equated by a single dust component, which is thinner and larger than the stellar component, as expected. We compare this equivalent dust component quantitatively with other studies in the next subsection. On the other hand, this result also implies that the projected image alone may not be sufficient to recover the detailed structure of the dust component. To reveal the three dimensional dust geometry of galaxies, a comprehensive study on the different dust attenuation properties is needed.

\subsubsection{Compare with other dust attenuation models}
\label{compare with other geometric models}

In this sub-section, we compare the dust geometry of our model galaxy with the result of other model galaxies. There are two widely used  dust geometric models of disk galaxies in literature, one is the single dust component model of  X99 and the other is the two dust component model of T04. 

To have a consistence comparison with the CCC model,  we set the stellar disk scale-length $R_s=1.5$ arcsec and the total un-attenuated stellar emission to  17.5 mag for both of the model galaxies in X99 and T04.

\subsubsubsection{X99 model}

The dust attenuation model of X99 is the same as the single dust component model we discussed in Section \ref{equivalent single dust component}, with the only difference being the inclusion of an additional bulge component in its stellar emission.

For the specific model parameters of X99, we adopt the typical values of their sample galaxies: $B/T\sim 0.2$, stellar scale-height to scale-length ratio $h_s/R_s\sim0.1$, dust to stellar scale-height ratio $h_d/h_s \sim 0.5$, dust to stellar scale-length ratio $R_d/R_s\sim 1.4$,V-band face-on optical depth $\tau_{f} \sim 0.5$.  We then make its V-band edge-on view image  and show it in the second top panel of Figure \ref{img}. The corresponding effective optical depth  and surface brightness profiles along the galactic plane  are shown as the solid curves in the bottom two panels.

As can be seen from the projected images, the X99   model shows a more pronounced  dust lane than CCC model. The effective dust optical depth along the galactic plane of the X99 model is systematically  larger than that of the CCC model ($\Delta {\tau}_{\rm eff} \sim 1$). However, for the surface brightness profiles, it is interesting to see that our CCC model prediction  is quite close to that of the typical galaxy in X99 ($\Delta \mu < 0.5 \rm{mag \cdot arcsec^{-2}}$ at all radii).

Before further discussion, it is worth reminding that the geometry of the stellar emission of the typical X99 model galaxies is different from that of the statistical MW-like galaxy in our CCC model.  First of all, the CCC model galaxy does not include a bulge component  whereas the X99 model galaxy has $B/T \sim 0.2$. Second, the stellar scale-height to scale-length ratio  of the CCC model galaxy ($h_s/R_s \sim 0.2$) is larger than that of the X99 model galaxy ($h_s/R_s \sim 0.1$). The reasons for these differences are twofold. On the one hand, the higher $h_s/R_s$ of the CCC model is partly a compensation for the absence of the bulge component in its model assumption (see more discussions in  Section \ref{Bulge component}). On the other hand, the $h_s/R_s$ of CCC model is fitted from a statistical sample of MW-like galaxies, while X99 model only fits 7 nearby edge-on galaxies. That is to say, the sample galaxies in these two models may not be comparable. 

The comparable $\mu$ profiles of two models are combined results of  different effects.  First, due to the higher $h_s$ of the CCC model, the un-attenuated surface brightness profile along the galactic plane of the CCC model is fainter.  On the other hand, the  scale-height of the equivalent single dust component of the CCC model galaxy ($h_d \sim 0.14 R_s$, Section \ref{equivalent single dust component}) is about twice  of that of the X99 model ($h_d \sim 0.07 R_s$), which makes effective optical depth  along the galactic plane to be significantly larger in the X99 model. Because of this significant large ${\tau}_{\rm eff}$, the stellar emission of the X99 model galaxy is more attenuated, so that it has a comparable and even fainter $\mu$ profile along the galactic plane. These complementary effects indicate the degeneracy between the stellar emission and the dust component in the modeling of the dust attenuation of galaxies. In other words, in the dust attenuation model, it is better not to predetermine the geometry of the stellar emission component, otherwise the geometric properties of the dust component obtained by modelling could be biased.

\subsubsubsection{T04 model}

The T04 study provides a framework of dust attenuation process with multi-components, where the stellar composition is composed of a bulge, a thick disk of old stars, and a thin disk of young stars, and the dust composition also includes both a thick and thin disk respectively.  The thick dust disk, which represents the continuously distributed ISM dust, is thinner and more extended than the old stellar disk, while the thin dust disk  has the same geometry as the young stars and represents the dust associated with new-born stars. Most of the geometric parameters of the T04 model use the values of the nearby galaxy  NGC 891  obtained from radiative transfer modelling (see T04 for detail). The only three free parameters excepted inclination remained in T04 are: the face-on optical depth $\tau_f$, bulge-to-total ratio B/T, and clumpiness $F$, where $F$ is defined as the total fraction of UV light being locally absorbed by dust in the thin disk. By giving different parameter settings for these three parameters, the T04 model is capable of describing the attenuation of galaxies at different inclinations and and therefore is widely used to study the attenuation-inclination dependence of disk galaxies\citep{Giessen2022,Driver2007,Masters2010}.

We also make the edge-on projection image for MW-like galaxies using the T04 model. Specifically, we first take the basic  geometric parameters from Table 1 of T04.  For the free parameters, we take the values from Table 3 of G22, where the T04 model parameters for the MW-like galaxies in SDSS have also been constrained from the stellar attenuation-inclination relation:  $B/T\sim 0.21$, $\tau_{\rm{V},f}=3.05$, and $F=0.34$ . Here, we convert the  $B$ band face-on dust optical depth $\tau_{\rm{B},f}$ to  $V$ band by dividing a factor of 1.32 (for $R_V=3.1$ extinction curve). The clumpiness factor $F$ is not related to the dust attenuation of old stellar population, and therefore  does not plays a role in the projection. 
The  edge-on view image projected from this T04-G22 model galaxy is shown at the top third panel  of Figure \ref{img}, whereas its ${\tau_{\rm eff}}$ and $\mu$ profiles are shown as the two dashed lines in the bottom two panels respectively.

As can be seen, the dust lane of this T04-G22 model galaxy is extremely prominent. As a result, its $\mu$ profile is significantly fainter than both of the  CCC and X99 model galaxies. Moreover, in this T04-G22 model galaxy, the $\mu$ profile is almost a constant out to $4$ times $R_s$, which means that the galactic plane is optically thick even to  its very out region.  This very optically thick surface brightness profile along the galactic plane (or the very prominent dust lane) of the T04-G22 model galaxy is a combined result of the  relatively large  dust content ($\tau_{\rm{V},f} \sim3$) and the very small height-to-length ratio ($0.016$)  of the thin dust disk preset in T04.  A detailed discussion of the origin of this atypical profile of the T04-G2 model galaxy is beyond the scope of this study. However, we will discuss more about the dust attenuation on emission lines  of the T04 model in Section \ref{HII regions}).

    \subsection{Optical Depth}
    \label{optical depth}

In the CCC model, besides the geometric parameters, we also have obtained constraints on the optical depth of two dust disks. 

For the diffuse dust disk, we get an estimate of the central absorption coefficient  $\alpha_{\rm{s,0}}\sim 1.22$ kpc$^{-1}$.  However,  this central parameter $\alpha_{\rm{s,0}}$ in our model is used more as a normalization parameter to describe the optical depth of the diffuse dust component at different regions rather than has an unambiguous physical implication of its own. The reason is that our exponential disk model is too simplified to describe the central region of a real galaxy that contains other complicate physical components, e.g., bulge, nuclear star cluster, and active galactic nuclei etc., which have not been taken into account. On the other hand,  it is worth using  $\alpha_{\rm{s,0}}$  to estimate the optical depth of typical regions of our the model galaxy. For example, at "solar neighborhood" (the location with a distance of $\sim$8.3 kpc from the model galaxy center on its galactic plane), our model predicts an absorption coefficient of $0.02$ kpc$^{-1}$. Then, by integration of the absorption coefficient along the line of sight to high galactic latitude regions, we obtain a line of sight optical depth  $\sim 0.01$ for diffuse ISM dust, which is in good consistence with the SFD map data of our MW \citep{Schlegel1998}.

For the clumpy nebular disk, we get  estimates of the optical depth $\tau_{\rm{cl}}\sim 0.5$ for each clump and the central absorption cross-section of clumpy regions $\sigma_{\rm{g,0}}\sim 0.84$ kpc$^{-1}$ . We remind that $\sigma_{\rm{g,0}}$ is the product of the central number density of the clumpy regions and projection area of each clump, $\rho_{\rm{g,0}}{\pi}R^2_{\rm{cl}}$.  For $\sigma_{\rm{g,0}}$, similar to the argument for the diffuse ISM dust, our estimate also can not be directly compared with the observations. However, we may use it to further probe the overall properties of the clumpy HII regions.  Before that, we need an estimate of the size of the individal clumpy region, $R_{\rm{cl}}$. The sizes of Galactic HII regions detected by their middle infrared (MIR) emissions are about $\sim 10$ pc \citep{Anderson2014}, while the sizes of giant HII regions detected in the nearby disk galaxies are shown in the range $10\sim 100$ pc in the optical wavelength \citep{Gutierrez2008}. The different sizes of HII regions detected in Galactic and extra-galactic disks may reflect the clumpy nature of the HII regions and possible selection effect \footnote{Because of the resolution effect, we expect that the observations of  HII regions in extra-galactic galaxies will be biased to larger ones. }.  Theoretical studies on the Str\"{o}mgren sphere also show that the size varies greatly with different H atom density and in different luminosity class of OB stars, ranging from several pc to nearly a hundred pc \citep{Gutierrez2008}. Here, for simplicity, we assume $R_{\rm{cl}}\sim 30$ pc and then obtain $\rho_{\rm{g,0}} \sim 300 $ kpc$^{-3}$.  With this  number, we  can further  obtain an  estimate of the total number of clumps inside the clumpy disk, $N=4{\pi}R_{\rm{g}}^2h_{\rm{g}}\rho_{\rm{g,0}}\sim 9.2\times10^{3}$, where we have adopted $R_{\rm{g}}=3.33$ kpc, $h_{\rm{g}}=0.22$ kpc in physical units.  If we assume that each clump has $ 10^3M_\odot$ newly formed stars (a HII region or open cluster usually contains $ \sim 10^3$ stars, initial mass function(IMF) also shows that only about $0.1\%$ stars are OB stars), we obtain a total of $\sim 10^{7}M_\odot$ new-born stars in our model galaxy. Considering the fact that the typical age of a HII region is about 10 Myr, and assuming that all new-born stars are formed inside the HII regions, the star formation rate of our model galaxy is then $SFR\sim 1 M_\odot $/yr, which is in excellent agreement with that of the local main-sequence star-forming galaxies \citep{Brinchmann2004}.  In  above discussion, we have used approximations for both $R_{\rm{cl}}$ and the number of newborn stars in each HII region. However, we would like to emphasize that our final estimate of the star formation rate does not significantly depend on the specific values of these two parameters. The reason is that these two parameters are physically correlated. If we assume a smaller $R_{\rm{cl}}$, the number of newborn stars in each HII region should also be smaller.

For the optical depth of clumpy regions, our model constraint, $\tau_{\rm{cl}}\sim 0.5$, is close to the typical value of the Galactic HII regions \citep[]{Sun2021}, but at the lower limit of the observed HII regions in  extra-galactic galaxies. \citep{Gutierrez2008}. Considering the clumpy nature of HII regions and the selection and resolution effects of observations, the HII regions observed in extra-galactic galaxies are more likely biased to giant HII regions or HII region groups. Indeed, as we have discussed, the sizes of HII regions reported in extra-galactic studies are systematically larger than that of Galactic ones.

With the CCC model, we may further have an estimate of the total amount of dust in each dust component. The total amount of dust in the diffuse ISM is
\begin{equation}
M_{\rm{d,s}}=\frac{\alpha_{\rm{s,0}}}{\kappa_V} 4{\pi} R_{\rm{s}}^2 h_{\rm{s}}=\frac{27.7 \mathrm{kpc}^2}{\kappa_V},
\end{equation}
where $\kappa_V$ is the mass absorption coefficient and in unit of  kpc$^2/\mathrm{M}_\odot$. For clumpy HII regions, the total dust amount is
\begin{equation}
\begin{aligned}
M_{\rm{d,g}}=&N_{\rm{cl}}M_{\rm{cl}}=\rho_{\rm{g,0}} 4{\pi} R_{\rm{g}}^2 h_{\rm{g}}\frac{4{\pi}R_{\rm{cl}}^3}{3}\frac{\tau_{\rm{cl}}}{R_{\rm{cl}}\kappa_V} \\
=&\frac{16{\pi} R_{\rm{g}}^2 h_{\rm{g}}\sigma_{\rm{g,0}}\tau_{cl}}{3\kappa_V}=\frac{17.1 \mathrm{kpc}^2}{\kappa_V}.
\end{aligned}
\end{equation}
Combining these two estimates, we conclude that, for MW-like galaxies, the total amount of dust in clumpy regions is less than  but comparable to that of the diffuse ISM dust.  Here, it is worth mentioning that we have not considered  the fully optically thick HII regions in above discussion (see more discussions in Section \ref{HII regions}).

\subsection{Attenuation Curves}
\label{attenuationcurves}

\begin{figure}[htbp]
    \centering
    \includegraphics[width=3in]{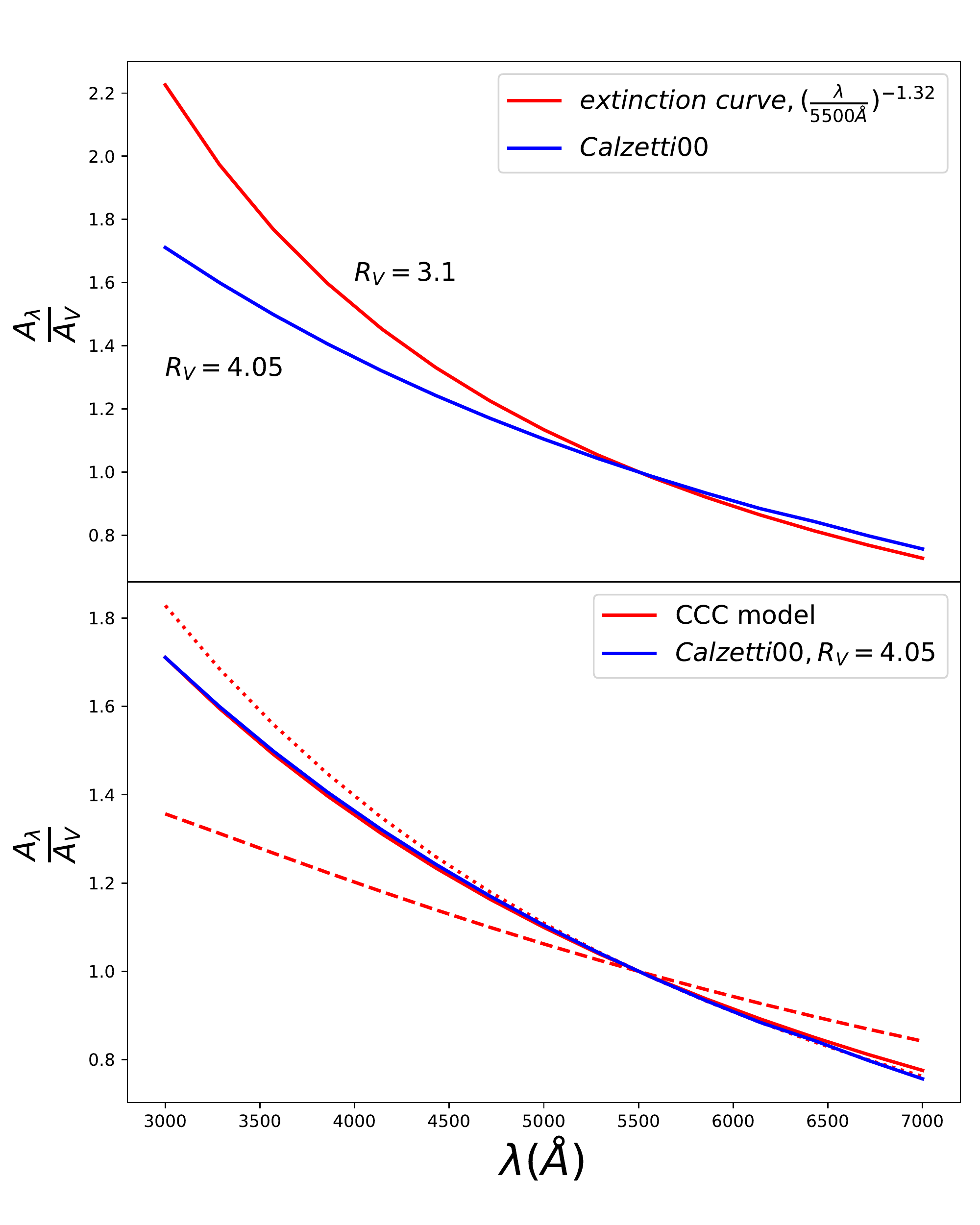}
    
    \caption{Extinction and attenuation curves of the CCC model and the Calzetti law represented by $\frac{A(\lambda)}{A_V}$. Top panel: the extinction curve (Equation \ref{power curve}) used in the CCC model (red curve) and the Calzetti attenuation curve (blue curve). Bottom panel: the attenuation curves derived from the CCC model for the edge-on case (dashed, $\theta = 90^{\circ}$, $R_V= 7.0$), face-on case (dotted, $\theta = 0^{\circ}$, $R_V= 3.7$) and median case (solid,$\theta = 60^{\circ}$, $R_V=4.1$), the blue curve also shows the  Calzetti attenuation curve for comparison.}
\label{attcurve}
 \end{figure}

In the CCC model, we have used a power law extinction curve (Equation \ref{power curve}) with $R_V=3.1$ for both the diffuse ISM dust and clumpy dust. However, during the SPS fitting to derive the $E(B-V)$ from the stellar continua of our sample galaxies, we used the Calzetti attenuation curve with $R_V=4.05$ \citep{Calzetti2000}. We plot these two different extinction (attenuation) curves on the top panel of Figure \ref{attcurve}, which are significantly different from each other at  short wavelengths ($\lambda <  5000\text{\AA}$). The difference is mainly due to the differences of their definitions, which we discuss in detail in following.

The extinction curve is only determined by the physical and chemical properties of the dust particles, while the attenuation curve, or the effective extinction curve, is further correlated with the geometrical distributions of both the dust particles and radiation sources \citep[e.g.,][]{Calzetti1997,Calzetti2000,Witt1992,Witt1996,Witt2000}. Since the CCC model gives a full description on the geometry of the dust particles and radiation sources, we can easily derive the shape of the attenuation curves for our model galaxy at different inclinations with the assumed power-law extinction curve ($R_V=3.1$).
In our modelling, we used $E_{\rm{g}}(\rm{H}\alpha-\rm{H}\beta)$ to represent the dust reddening of the nebular emission (Equation \ref{Dexp}), which is independent of the extinction curve being assumed. Therefore, we only need to consider the shape of the attenuation curves for stellar continua.

We use the CCC model and take the best estimates of the model parameters (Table \ref{2com}) to calculate the effective dust attenuation for the stellar continua of our model galaxy at different wavelengths and then derive the shape of the attenuation curve. We show the resulted attenuation curves for three representative inclination angles,  $\theta = 0^\circ, 60^\circ$, and $90^\circ$, as the red dotted, solid and dashed lines in the bottom panel of Figure \ref{attcurve}, which have $R_V= 3.7, 7.0,$ and $4.1$, respectively.  We see the CCC model naturally predicts an increasing of $R_V$ with increasing disk inclination, which has been reported in observational studies \citep[e.g.][]{Battisti2017}. The increasing of $R_V$ of the attenuation curve with disk inclination (optical depth) is caused by the saturation effect when the emission sources are mixed with the dust (see Equations \ref{mix_tau} and \ref{mix_e}). Also, for the median disk inclination ($\theta\sim60^\circ$), the CCC model predicts an attenuation curve with $R_V=4.1$, which is in excellent agreement with the classical Calzetti attenuation curve \citep{Calzetti2000} with $R_V=4.05$. Moreover, the ranges of the attenuation curves $R_V \in (3.7, 7.0)$ predicted from the CCC model  is also matched with the observational results of nearby galaxies \citep{Calzetti1997}.

In Section \ref{sec:es}, we used the Calzetti attenuation curve with an constant $R_V=4.05$ in the SPS fitting. To make the picture fully self-consistent, we should apply different attenuation curves for galaxies with different inclinations. However, introducing the inclination effect comprehensively into the SPS fitting process is beyond the scope of this work. Nevertheless, we expect that this inconsistency does not have a significant impact on our conclusions.  For example, the derived median color excess $E_{\rm{s}}(B-V)$ of a sample of nearby galaxies only changes from 0.15 to 0.16 when the attenuation curve is changed from $R_V=4.05$ to 4.88 \citep{Calzetti1997,Calzetti2000}.

\subsection{Caveats}
\label{caveats}

Our CCC model has not only presented excellent fits to the complicate inclination dependence of both the nebular and continuum reddening features(Figure \ref{CCC_model}), but also give  consistent predictions on their dust attenuation effects (Figure \ref{apcorr}).  However, our model is also subject to limited  observational constraints and  uncertainties in the model assumptions. We discuss these caveats in the CCC model below.

\subsubsection{Bulge component}
\label{Bulge component}

For simplicity, our CCC model has not considered the bulge component of disk galaxies. A reasonable assumption is that the bulge component does not have star formation, so there is no cold gas and dust associated. In this case, the dust attenuation of the nebular lines $E_{\rm{g}}$ would be independent of whether or not including a bulge component in our model. For stellar emission, if we assume that the bulge component is concentrated and spherical,  then the combination of a spherical bulge and a thin disk is essentially equivalent to a slightly thicker disk. This is precisely the reason for the relatively high stellar disk scale-height to length ratio,  $h_s/R_s \sim 0.19$，obtained by our CCC model. A more refined geometric model with a bulge component introduces more free parameters, and therefore requires more observational constraints, e.g. the combination of multi-wavelength images, and/or sub-samples of galaxies with similar bulge-to-disk ratios.

\subsubsection{dense clouds}
\label{HII regions}

In our CCC model, the HII regions have been assumed to be identical. The best fit of the optical depth,  $\tau_{\rm{cl}}\sim 0.5$,  can be considered as a statistical average of different HII regions.  However, in real galaxies,  the optical depths of star forming regions are related to their evolution phase \citep{McKee2007}. The star forming regions at their early stage, which are still embedded in molecular clouds, could be extremely dense and optically thick and  make no contribution to the observed optical nebular emission lines at all. Not only that, these optically thick star forming regions would also block the stellar emission behind them and thus cause a significant fraction of dark area when galaxies are viewed from edge-on.

Our modelling is based on the observational constraints from the Balmer decrement of the optically thin HII regions and the reddening of stellar continuum in optical wavelengths. Therefore, these optically thick star forming regions will not directly bias our model results. Moreover, our results in Figure \ref{apcorr} also show that the CCC model predicted dust attenuation effects are in good consistence with observations once the aperture effects have been probably accounted. That is to say, at least inside the SDSS fiber aperture, there will not be many of these completely optically thick HII regions that could result in  significant  areas being completely obscured. However, we cannot exclude that, when the galaxies are completely edge-on, these dark areas will have a non-negligible effect on the projected image. Indeed, the dust optical depth along the galactic plane predicted by our CCC model is slightly smaller than that of the X99 model, which is obtained from RT modelling and should be unaffected by these dark clouds. (Section \ref{Dust geometry}).

The optically thick star forming regions have been properly accounted in T04 model with a parameter $F$ (volume fraction of optically thick HII regions), which has a typical value of 0.2, but varying significantly in different galaxies. As the setup in the T04 model, these optically thick star forming regions would have a significant impact on the UV radiation from very young stellar objects and the corresponding  transferred infrared emission.  For future studies, these optically thick components need to be taken into account, especially when there are observational constraints from either the UV or IR.

\subsubsection{Resolution of fiber spectroscopy }
\label{resolution of fiber spectroscopy}

In this study, we mainly model the dust attenuation effects along the line of sight to galactic center since our observational constraints are mainly from the fiber spectroscopy of sample galaxies. As we have discussed in Section \ref{Off-axis effect} and \ref{APERTURE EFFECT AND DUST ATTENUATION}, the fiber aperture, although much smaller than typical sample galaxy size and can approximate the galaxy central region well for face-on galaxies, which also brings significant biases on interpreting the observed dust attenuation features of edge-on galaxies.

Here, we present another bias effect from the resolution of fiber aperture  that has not been discussed yet. For edge-on galaxies, because the galaxy scale-height is much smaller than the fiber spectroscopy, the observed dust attenuation is not simply along the galactic plane  but rather an integral along the vertical direction of the disk. As  shown by  the projected images of edge-on galaxies, the optical depth decreases significantly from the galactic plane to high latitudes, which thus makes the observed dust attenuation  be systematically smaller than that completely along the galactic plane. In our modeling, for edge-on galaxies, we have only counted the dust optical depth and calculated the dust attenuation effect on the galactic plane. That is to say, for edge-on galaxies, our model fitting values are biased towards lower dust attenuation, which would bias the clumpy dust disk to be relatively thick. Indeed, as we have shown  in Figure \ref{img}, the effective optical depth along the galactic plane predicted from our CCC model is smaller than that of the X99 model.   However, the observational constraints of the CCC model are the reddening features of  all inclinations (Figure \ref{egest}, 90 $\theta$ bins), we therefore do not expect such a bias from edge-on galaxies only could significantly change our model results.

\subsubsection{IR properties}
\label{IR properties}

In this study, our model has only investigated the dust  absorption process without considering its emission. To better discuss and constrain the dust properties of galaxies, a  radiative transfer model that takes into account  IR emissions of dust is required, which however is beyond the scope of  current work. Here, based on the framework and basic results of our CCC model, we give a brief outlook on the infrared emissions of our model galaxy.

In our CCC model, we have assumed that the dust in both clumps and ISM have the same properties (extinction curves), thus the interstellar radiation field (ISRF) plays the key role in dust temperature. Considering that the clumpy HII regions are radiated by central young stellar objects, we expect that the clumpy dust  will absorb more UV photons and therefore  constitute the warm dust that could be traced by  MIR emissions.  Indeed, in observation, the warm dust is spatially correlated with the molecular gas $\rm{H}_2$ and star forming regions \citep{Stevens2005,Hippelein2003}. It is also shown that the scale-length  of the MIR disk is similar to that of $\rm{H}\alpha$ \citep{Vogler2005}.

On the other hand, our CCC model shows that the clumpy dust is more extended than that of diffuse dust. If we assume that the clumpy dust in the HII region has averagely higher temperature than that of ISM dust, the disk will be more extended in the MIR than in the FIR, which is on the contrary to the observations that the disk scale-length of the IR emission increases with wavelength \citep{Hippelein2003}. One of the reasons for this contradiction is that our model does not contain any optically thick clumpy components, as discussed in Section \ref{HII regions},  which has no impact on the model fitting since all our modelling constraints are in the optical wavelengths. These optically thick star-forming regions (e.g., molecular clouds)  are one of the main contributors of cold dust \citep{PlanckXXV2011}. If we take the IR properties of these optically thick star-forming regions into account and assume that they have the same geometrical distribution as the HII regions in CCC model, we would naturally get a more extended cold dust component. Another reason is that a uniform and low dust temperature assumption for the diffuse ISM dust is oversimplified. The temperature of the diffuse ISM dust is positively correlated with the intensity of ISRF, which decreases from the galactic center to the outer regions. Therefore, the spatial distribution of the lower temperature component of diffuse ISM dust will also be biased to a larger scale-length. To fully quantify the IR emissions of our CCC model, we need to model the radiative transfer of both the clumpy dust (optically thin HII regions and optically thick molecular clouds) and the diffuse ISM dust in detail, which is beyond the scope of this work and requires further investigation in future works.

\section{CONCLUSION}
\label{CONCLUSION}
In this study, we have measured the dust  reddening features from the fiber spectra of a sample of 33,065 MW-like  disk galaxies in the SDSS, where the stellar reddening is determined using the full-spectrum SPS code STARLIGHT \citep{Fernandes2005}, and the nebular emission line reddening is measured from the Balmer decrement. We explore the variation of these two different dust attenuation tracers as a function of disk inclination and then use it as a constraint to build geometric models for the dust attenuation of MW-like disk galaxies. 

We find that the stellar attenuation shows a monotonic increase with disk inclination while the nebular attenuation does not. For highly inclined disks ($\theta>75^\circ$), although the fluxes of emission lines (e.g. $\rm{H}\alpha$) continue to decrease with increasing of disk inclination, the nebular attenuation (reddening) shows a saturation effect: $E_{\rm{g}}(B-V)_{max}\sim 0.6$.  

For the model part, which is  also the focus of this study, we find that a single uniform mixture model can generally reproduce the observed inclination dependence of the stellar attenuation, while a single screen model can only partly reconstruct the inclination dependence of the nebular attenuation for low inclination disks. \textit{Based on the results of these two simple models, we construct a new two-component dust geometry model, the Chocolate Chips Cookie  (CCC) model. In the CCC model, the clumpy nebular regions are embedded in a diffuse stellar/ISM disk, like chocolate chips in cookies.} By adopting a clumpy approximation and considering the off-axis effect of highly inclined disks, the CCC model successfully reproduces the observed inclination dependence of the stellar and nebular reddening simultaneously. Moreover, after proper accounting for the fiber aperture effect, the CCC model prediction on the inclination dependence of the dust attenuation effects of both $\rm{H}\alpha$ flux and $r$ band magnitude are also in good consistence with observations. In addition to the observational properties from fiber spectroscopy,  the global edge-on photometric properties predicted  from our CCC model are also broadly consistent with the RT studies for nearby disk galaxies.

The best estimates of the geometric parameters of our model galaxies are as follows:  the stellar disk scale-height to scale-length ratio $h_{\rm{s}}/R_{\rm{s}}\sim 0.19$, nebular disk scale-height to scale-length ratio $h_{\rm{g}}/R_{\rm{g}}\sim 0.06$, the ratio between the nebular disk scale-height to diffuse disk $h_{\rm{g}}/h_{\rm{s}} \sim 0.56$, and the ratio between the nebular disk scale-length to the diffuse disk $R_{\rm{g}}/R_{\rm{s}} \sim 1.6$. Moreover, we obtain model constraints on the optical depth of two dust disks. For the diffuse dust component, the CCC model predicts a line of sight dust reddening to high galactic latitude at "solar neighborhood" $E(B-V)\sim 0.02$. This result indicates that statistically, our MW-like disk galaxies selected from the SDSS according to their stellar mass have similar geometry and dust properties to the MW as well. For the clumpy regions, we conclude that there are about $ 10^4$ clumpy regions in our model galaxy if taking 30 pc as the clump size, and each region has an optical depth $\tau_{\rm{cl}}\sim 0.50$ in $V$ band. These parameter estimates give a self-consistent inference that these clumpy regions can properly represent the HII regions in SFGs.

The CCC model framework also has some limitations. First, in order to reduce the number of free parameters, we have not considered the bulge component, which makes the scale-height of the stellar emission $R_s$ in the CCC model is an effective parameter,  containing the bulge contribution. Secondly, our model has not considered  the optically thick star forming regions, which may prevent the direct application of our model to radiative transfer studies involving UV and/or IR data.  Finally, our CCC model is developed to study the dust attenuation process along a single line of sight, which poses difficulties in detailed comparison of our model results with  the  finite resolution observational data in a straightforward way.

The CCC model has various applications. For example, our model can be further extended and applied to the disk galaxies with other stellar masses. With such modeling, the structural parameters of the dust components of the galaxies with different masses can be compared and explored, enabling us  a better understanding of the formation and evolution of the dust components in galaxies with different masses. Moreover, by assuming a physical prescription of the properties of the dust particles \citep[e.g.][]{Draine2007}, our model can be applied on more detailed dust extinction and emission processes in different wave-lengths, which then can be compared with the results from numerical simulations using Monte-Carlo processes \citep[e.g.][]{SKIRT}.  The CCC model is an analytical model based on several reasonable approximations. It simplifies the calculation of radiative transfer process when considering the star-dust geometry. Comparing with the common Monte-Carlo method for calculating the detailed radiative transfer process, the CCC model is much faster, and thus can be applied for large samples of galaxies conveniently.

\section*{acknowledgments}

We thank the anonymous referee for the helpful comments that improve the paper. This work is supported by  the National Natural Science Foundation of China (No. 12073059 $\&$ No. U2031139) , the National Key R$\&$D Program of China (No. 2019YFA0405501) . We also acknowledge the science research grants from the China Manned Space Project with NO. CMS-CSST-2021-A04, CMS-CSST-2021-A07, CMS-CSST-2021-A08, CMS-CSST-2021-A09, CMS-CSST-2021-B04. F.T.Y. acknowledges support by the Funds for Key Programs of Shanghai Astronomical Observatory (No. E195121009) and the Natural Science Foundation of Shanghai (Project Number: 21ZR1474300).

Funding for the SDSS and SDSS-II has been provided by the Alfred P. Sloan Foundation, the Participating Institutions, the National Science Foundation, the U.S. Department of Energy, the National Aeronautics and Space Administration, the Japanese Monbukagakusho, the Max Planck Society, and the Higher Education Funding Council for England. The SDSS Web Site is http://www.sdss.org/.

The SDSS is managed by the Astrophysical Research Consortium for the Participating Institutions. The Participating Institutions are the American Museum of Natural History, Astrophysical Institute Potsdam, University of Basel, University of Cambridge, Case Western Reserve University, University of Chicago, Drexel University, Fermilab, the Institute for Advanced Study, the Japan Participation Group, Johns Hopkins University, the Joint Institute for Nuclear Astrophysics, the Kavli Institute for Particle Astrophysics and Cosmology, the Korean Scientist Group, the Chinese Academy of Sciences (LAMOST), Los Alamos National Laboratory, the Max-Planck-Institute for Astronomy (MPIA), the Max-Planck-Institute for Astrophysics (MPA), New Mexico State University, Ohio State University, University of Pittsburgh, University of Portsmouth, Princeton University, the United States Naval Observatory, and the University of Washington.

\appendix
\renewcommand\thefigure{\Alph{section}\arabic{figure}}   

\section{The stellar age and metallicity of sample galaxies from SPS fitting}
\label{sec:z-m-a}
\setcounter{figure}{0}  
During stellar population synthesis (SPS) fitting, there is a well-known age-metallicity-attenuation degeneracy effect.  We use this effect to check the possible biases in our measurement of the stellar reddening $E(B-V)$. If $E(B-V)$ were biased, and since  $E(B-V)$ is a strong function of disk inclination, we would expect the disk inclination to be related to either the average stellar age or metallicity, or both. 

\begin{figure}[htbp]
    \centering
\gridline{\fig{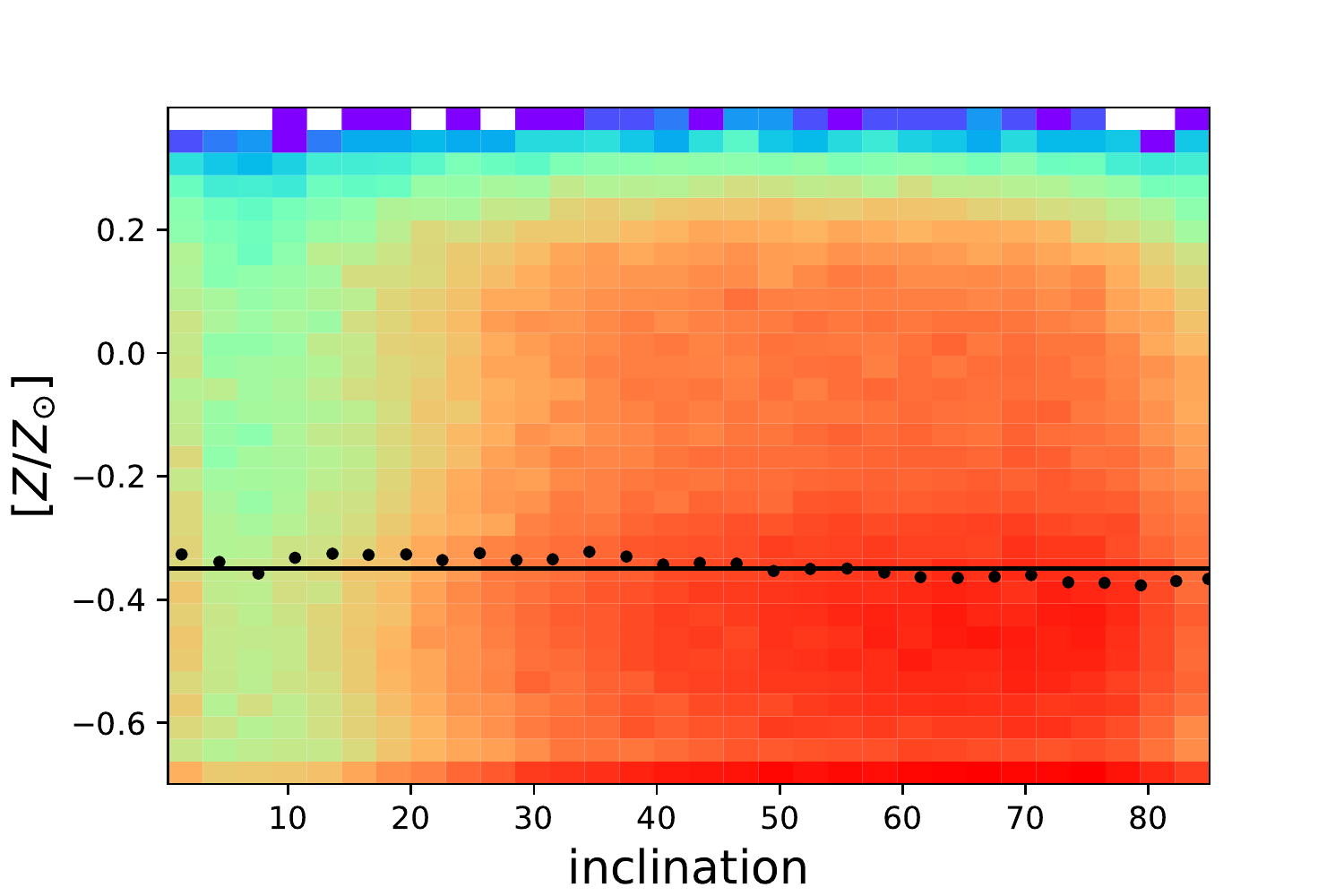}{2.5in}{(a) }
          \fig{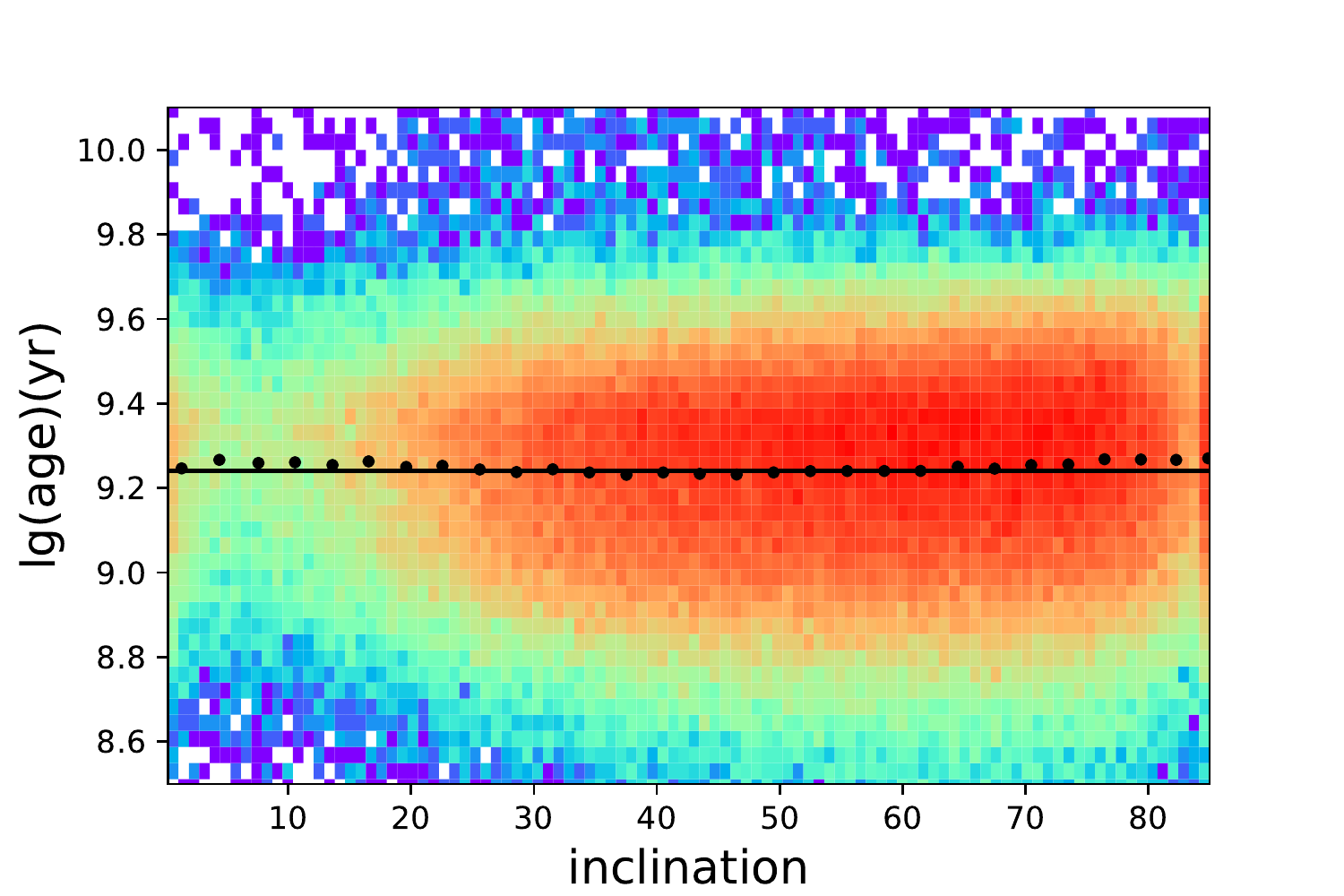}{2.5in}{(b) }
          }

    \caption{ The inclination dependence of average metallicity (left panel) and stellar age (righ panel)  obtained from SPS fitting of our MW-like galaxy sample.
    }
    \label{degenerate}
\end{figure}

We show the density map of the SPS fitting derived  average metallicity and stellar age of our MW-like SFGs (Section \ref{sec:es}) as function of  their disk inclination in Fig. \ref{degenerate}, where the median values at given disk inclination are shown as solid dots. As can be seen, there is no inclination dependence of both the age and  metallicity. 

\section{dust attenuation of clumpy clouds: Gaussian approximation}
\label{continuum approximation}
\setcounter{figure}{0}  
\setcounter{equation}{0}  

\begin{figure}[htbp]
    \centering
\gridline{\fig{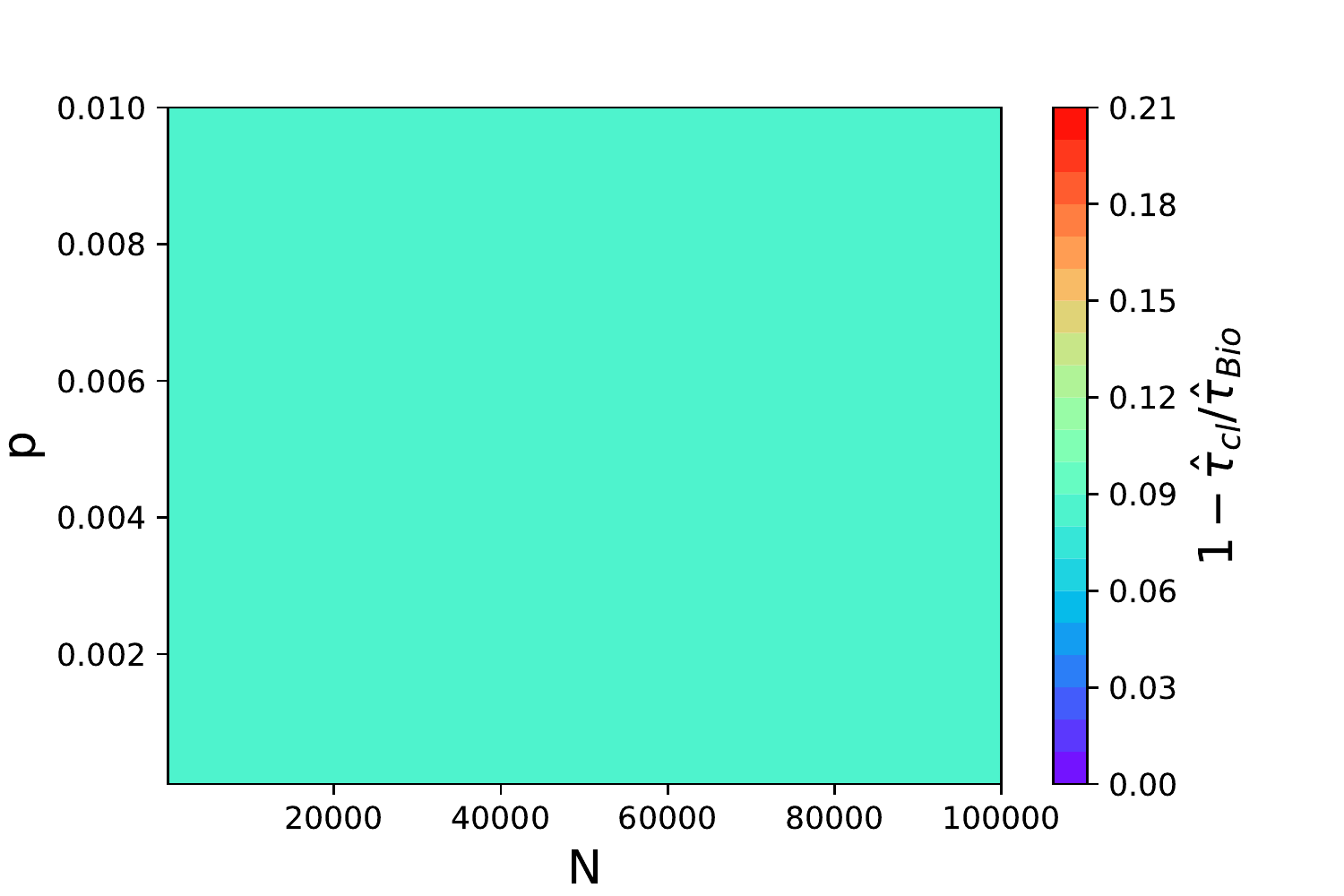}{2.5in}{(a) approximate deviation for $n$ and $p$}
          \fig{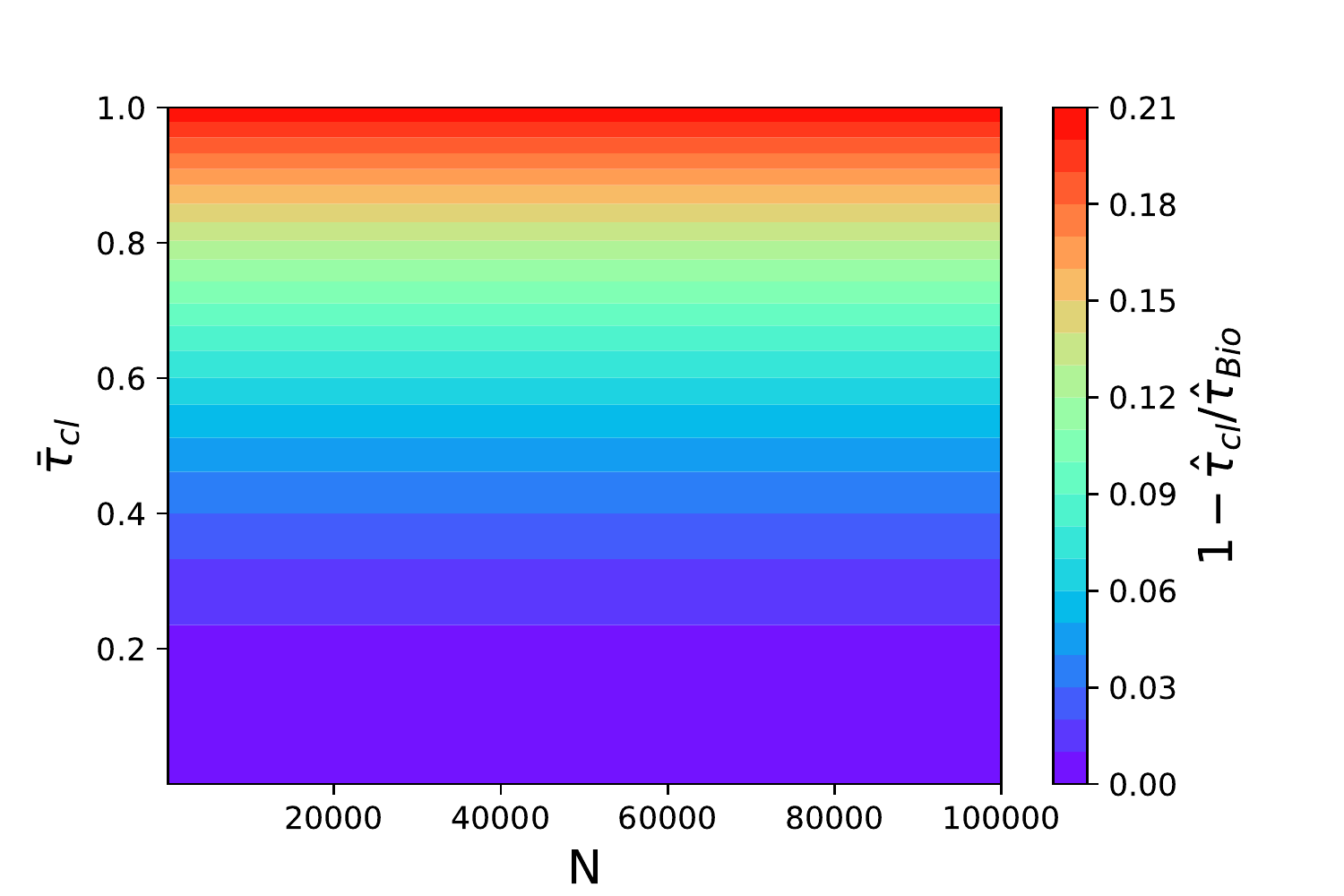}{2.5in}{(b) approximate deviation for $n$ and $\tau$}
          \fig{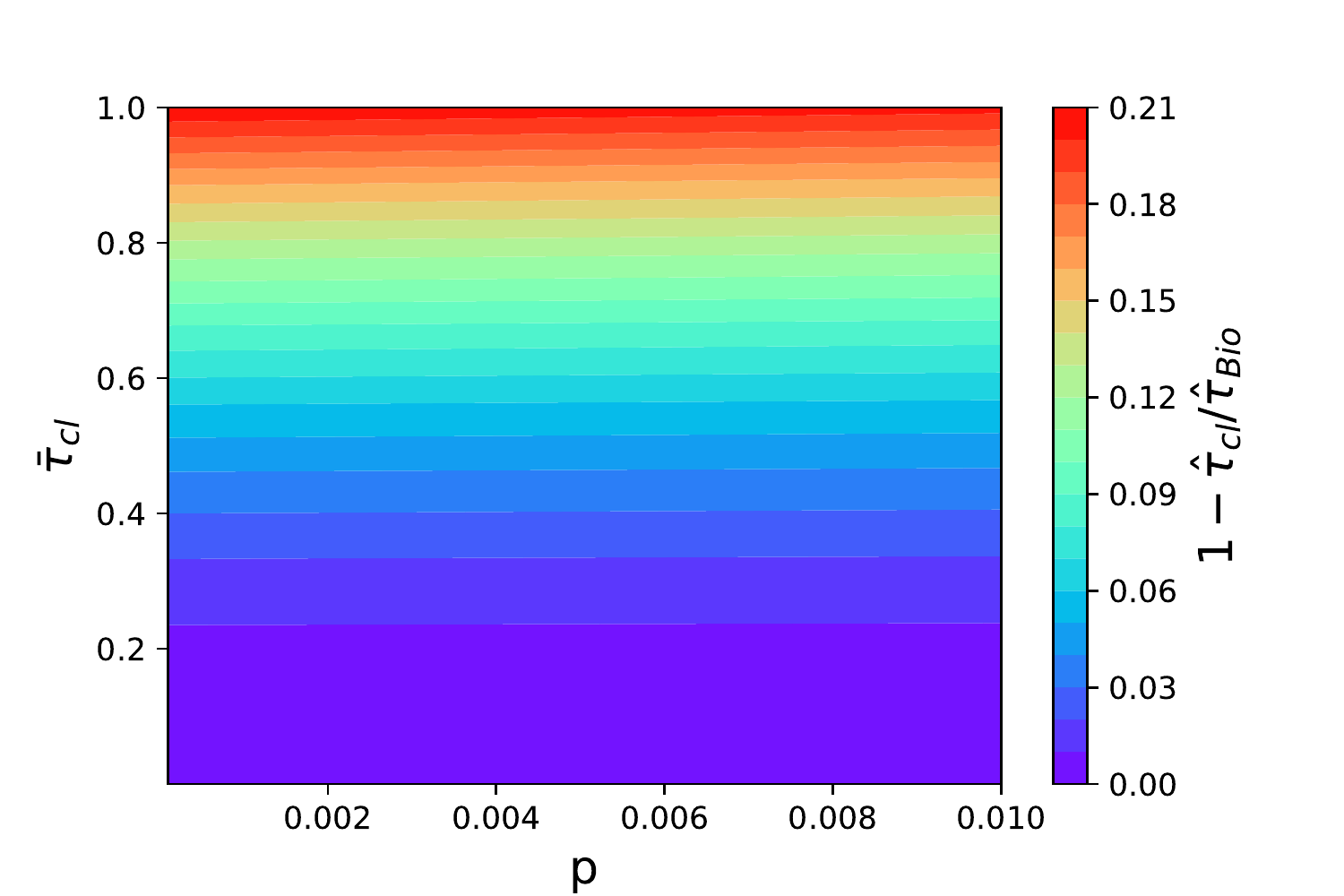}{2.5in}{(c) approximate deviation for $p$ and $\tau$}
          }

    \caption{The consistence of the model predicted optical depth of clumpy regions between the Gaussian approximation  and numerical calculation of binomial distributions  for different parameter sets, which are color-coded by $1-\hat{\tau}_{\rm{cl}}/\hat{\tau}_{\rm{Bio}}$ in each panel.  Panel (a):  $\bar{\tau}_{\rm{cl}}=0.67$,  $p$ ranges from 0 to 0.01 and $n$ ranges from 100 to 100,000; Panel (b):  $p=10^{-4}$, $n$ ranges from 10 to 100,000 and $\bar{\tau}_{\rm{cl}}$ ranges from 0 to 1; Panel (c): $n=10,000$, $p$ ranges from 0 to 0.01 and $\bar{\tau}_{\rm{cl}}$  ranges from 0 to 1.
    }
    \label{nptapprox}
 \end{figure}
 
Different from the classical continuous dust disk, we adopt a disk with discrete clumps where the optical depth along line of sight is written as 
\begin{equation}
\hat{\tau}_{\rm{Bio}}=-\ln\sum^N_{k=0}B(k|N,p)*\exp(-k\bar{\tau}_{\rm{cl}})\,
\label{taubio}
\end{equation}
which is then approximated with  Equation \ref{equtau}.

In specific, we have approximated the Binomial distribution $B(N,p)$ with a Gaussian distribution $\mathcal{N}(Np, Np(1-p))$ under the assumptions of  $N >> 1$ and $p<<1$.  However, due to the extinction term $\exp(-k\bar{\tau}_{\rm{cl}})$ in Equation \ref{taubio},  a very large optical depth may lead to deviations from this approximation. In this appendix, we make detailed comparisons between the numerical calculation of the binomial distribution extinction (Equation \ref{taubio}) and our Gaussian model approximation(Equation \ref{equtau}).

 For our numerical calculations, we examine the deviation between Equation \ref{equtau} and Equation  \ref{taubio} in terms of $1-\hat{\tau}_{\rm{cl}}/\hat{\tau}_{\rm{Bio}}$.   There are 3 parameters in Equation  \ref{taubio}: number of clumps $N$, covering factor $p$ and effective optical depth $\bar{\tau}_{\rm{cl}}$ of individual clump.  We show the deviation as functions of different sets of $N,p,\bar{\tau}_{\rm{cl}}$ in three different panels of Figure \ref{nptapprox}. In top, middle and bottom panel, we set  $n=10,000$, $p=10^{-4}$, $\bar{\tau}_{\rm{cl}}=0.67$ (the best estimates of the CCC model, $\bar{\tau}_{\rm{cl}}=4/3 \tau_{\rm{cl}}$) and let the other two parameters being free respectively.  The ranges  of these free parameters are  set as follows: 0 to 0.01 for $p$, 100 to 100,000 for  $N$ and 0 to 1 for $\bar{\tau}_{\rm{cl}}$.

 As can be seen,  inside the parameter range, the deviation is almost independent of $N$ and $p$, but significantly dependent on $\bar{\tau}_{\rm{cl}}$. For the best estimate of the CCC model, $\bar{\tau}_{\rm{cl}}\sim0.67$, the deviation is about $8\%$. However, we note that when  $\bar{\tau}_{\rm{cl}}\sim 1.0$, the deviation of the approximation Equation \ref{equtau} can be up to $20\%$. Moreover, this bias is a negative bias, i.e. it underestimates the effective optical depth of the clumps.
 
 \section{dust attenuation of clumpy clouds: difference from continuous distribution}
 \label{sec:diff of cont and cl}
 \setcounter{figure}{0}  
 \setcounter{equation}{0}  
 We present a quantitative comparison  between the dust attenuation of two different models: the clumpy distributed dust and  continuously distributed dust.   In specific, we consider two  models that both have the same amount of dust inside a given column $V=S*l$ (much larger than the size of dust cloud itself), but one is continuously distributed and the other is clumpy distributed. We illustrate the clumpy distribution and the continuous distribution in Figure \ref{cl_cont}.
 
 \begin{figure}[htbp]
    \centering
    \includegraphics[width=5in]{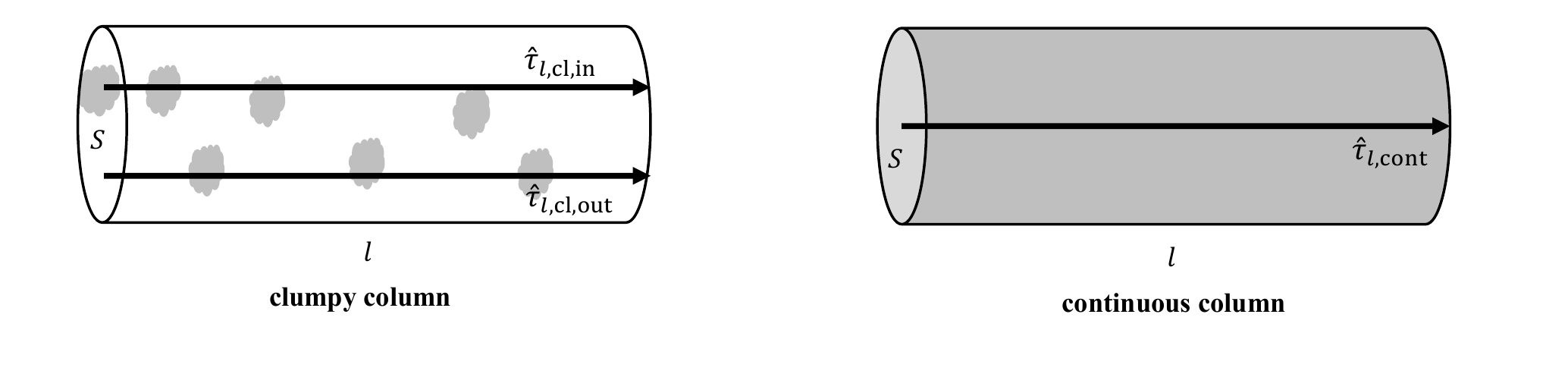}
    
   \caption{Cartoon of  clumpy dust distribution(left) and continuous dust distribution(right) for the same amount of dust in the same volume $V=S*l$.}
\label{cl_cont}
 \end{figure}
 
For the clumpy dust distribution, if there are $N$  dust clouds uniformly distributed inside $V$, the total amount of dust  is 
\begin{equation}
    M_{\rm{tot}}=\frac{4N\pi R_{\rm{cl}}^2 \tau_{\rm{cl}}}{3\kappa},
\end{equation}
where $R_{\rm{cl}}$ is the radius of the dust cloud, $\tau_{\rm{cl}}$ is the optical depth of a emission source at the center of the cloud, and $\kappa$ is the dust extinction coefficient. In this case, as we have shown in Section \ref{Modelling of the clumpy regions} (Equation \ref{equtau1}), the line of sight dust optical depths for a region outside cloud  and a region inside cloud can  be approximated by

\begin{equation}
 \hat{\tau}_{l,\rm{cl,out}}=\frac{n_{\rm{g}}{\pi}R^2_{\rm{cl}}\bar{\tau}_{\rm{cl}}(2-\bar{\tau}_{\rm{cl}})}{2}  \,,
\end{equation}
and
\begin{equation}
 \hat{\tau}_{l,\rm{cl,in}}=\tau_{\rm{cl}}+ \frac{n_{\rm{g}}{\pi}R^2_{\rm{cl}}\bar{\tau}_{\rm{cl}}(2-\bar{\tau}_{\rm{cl}})}{2}  \,,
\end{equation}

where $\bar{\tau}_{\rm{cl}}=\frac{4\tau_{\rm{cl}}}{3}$, and $n_{\rm{g}}=N/S$ is  the column number density of the dust clouds along the line of sight.

For the continuous dust distribution, the optical depth along the same line of sight can be expressed as

\begin{equation}
     \hat{\tau}_{l,\rm{cont}}=\frac{M_{\rm{tot}}}{S}\kappa=n_{\rm{g}}{\pi}R^2_{\rm{cl}}\bar{\tau}_{\rm{cl}} \ .
\end{equation}

As can be seen, there are two major differences between these two models. On the one hand, for the clumpy distributed dust,
the dust attenuation of the inside cloud regions and outside cloud regions are different with a term of $\tau_{\rm{cl}}$, while the continuous dust does not have this effect. On the other hand, for the diffuse ISM region, the clumpy distributed dust has a systematically smaller effective dust extinction than the continuously distributed dust, with a factor of $(2-{\bar{\tau}})/2$, which is also known as `mega-grain' effect \citep{varosi1999}. 

Combining these two effects, the attenuation of the emission lines by the clumpy dust  can be either greater or smaller than  the case of continuous dust, depending on the column number density of the clumpy clouds along the line of sight. For example, if we take $\tau_{\rm{cl}} = 0.5$ and assume $n_{\rm{g}} \pi R_{\rm{cl}}^2 = 0.5$,  corresponding to the face-on view of a typical disk galaxy, the resulted $\hat{\tau}_{l,\rm{cl,in}}$ is larger than $\hat{\tau}_{l,\rm{cont}}$ with a factor of $0.38$. If we take $n_{\rm{g}} \pi R_{\rm{cl}}^2= 4$ (approximating edge-on view), $\hat{\tau}_{l,\rm{cl,in}}$, on the contrary, is smaller than $\hat{\tau}_{l,\rm{cont}}$ with a factor of 0.5.

\bibliography{ref}
\bibliographystyle{aasjournal}

\end{CJK*}
\end{document}